\documentclass[a4paper,10pt]{article}

\usepackage{lscape}
\usepackage{cancel}
\usepackage{bm}
\usepackage{bbm}
\usepackage{multirow} 
\usepackage{amsfonts}
\usepackage{amssymb}
\usepackage{graphicx}
\usepackage{amsmath}
\usepackage[usenames,dvipsnames]{color}
\usepackage{hyperref}
\usepackage{epstopdf,epsfig}
\usepackage{color}

\usepackage[left=.7in,right=.7in,top=.8in,bottom=.8in]{geometry}             

\hypersetup{
    colorlinks=true,         
    linkcolor=blue,          
    citecolor=red,        
    urlcolor=Violet             
}

\newtheorem{defi}{Definition}
\newtheorem{theo}{Theorem}

\newcommand{\diby}[2]{\ensuremath{\frac{\delta #1}{\delta #2}}}

\newcommand{\order}[1]{\ensuremath{\mathcal{O}(#1)}}
\newtheorem{conj}{Conjecture}

\def\be{\begin{equation}}
\def\ee{\end{equation}}
\def\bea{\begin{eqnarray}}
\def\eea{\end{eqnarray}}

\title{Timeless configuration space and the emergence of classical behavior}
\author{\bf Henrique Gomes\footnote{\href{mailto:gomes.ha@gmail.com}{gomes.ha@gmail.com}}\\\it Perimeter Institute for Theoretical Physics\\ \it 31 Caroline Street, ON, N2L 2Y5, Canada}
\begin{document}
\maketitle

\begin{abstract}
The inherent difficulty in talking about quantum decoherence in the context of quantum cosmology is that decoherence requires subsystems, and cosmology is the study of the whole Universe. Consistent histories gave a possible answer to this conundrum, by phrasing decoherence as loss of interference between alternative histories of closed systems. When one can apply Boolean logic to a set of histories, it is deemed \lq{}consistent\rq{}. However, the vast majority of the sets of histories that are merely consistent are blatantly nonclassical in other respects, and further constraints than just consistency need to be invoked. In this paper, I attempt to give an alternative answer to the issues faced by consistent histories, by exploring a timeless interpretation of quantum mechanics of closed systems. This is done solely in terms of path integrals in non-relativistic, timeless, configuration space. What prompts a fresh look at such foundational problems in this context is the advent of multiple gravitational models in which Lorentz symmetry is not fundamental, but only emergent. And what allows this approach to overcome previous barriers to a timeless, conditional probabilities interpretation of quantum mechanics is the new notion of records -- made possible by an inherent asymmetry of configuration space. I outline and explore consequences of this approach for foundational issues of quantum mechanics, such as the natural emergence of the Born rule, conservation of probabilities,  and the Sleeping Beauty paradox.
\end{abstract}

\section{Introduction}

\subsection{Motivation}
The physical process of quantum decoherence began its life as an answer to the measurement problem. The role it plays in diagonalizing density matrices, however, is not enough for the satisfaction of the more stringent among us. For example, questions about wavefunction collapse persist to this day, but their actual importance is still up for debate. As Steven Weinberg put it: 
\begin{quote}
``So irrelevant is the philosophy of quantum mechanics to its use, that one begins to suspect that all the deep questions about the meaning of measurement are really empty, forced on us by our language, a language that evolved in a world governed very nearly by classical physics. But I admit to some discomfort in working all my life in a theoretical framework that no one fully understands." -  \cite{Weinberg_dreams} 
\end{quote}
Perhaps not enough time has yet passed to accommodate our intuition towards quantum mechanics, or perhaps there is something inherently ill-defined with our present understanding.  For whichever reason, echoing Weinberg, many researchers deem it premature to state that all foundational issues with quantum mechanics are resolved. The source of contention can be traced back to the different processes required in quantum mechanics: one is unitary evolution, the other a non-unitary reduction of the wave-function. 

\paragraph*{Collapse vs. unitary evolution.}
Many attempts exist to blur this distinction, decoherence being the most well-known physical mechanism entrusted with that responsibility. Apart from  debatable criticisms  having to do with how decoherence itself encodes the measurement process  (e.g. how the Born rule emerges from decoherence in the Many Worlds interpretation), the approach has one non-controversial objective shortcoming: it requires a separation between subsystems (or subsystems and an environment; the environment corresponds to the \lq\lq{}inaccessible\rq\rq{} degrees of freedom).\footnote{For the most up to date achievements of \lq{}environment-induced-decoherence\rq{}, see \cite{Zurek_envariance}.} In particular, it seems to be of limited use for quantum cosmology and quantum gravity. 

Perhaps, then,  a full resolution requires us to first apply some of the lessons learned from quantum gravity. One of these lessons is that \lq\lq{}Time\rq\rq{} --- a concept inextricably linked with both unitary evolution and non-unitary collapse--- is a more problematic concept than might first appear (see e.g. \cite{Kuchar_Time} for a review of the many issues that crop up). This problem is reflected in the standard Wheeler-DeWitt equation, $\hat{H}|\Psi(\phi)\rangle=0$, a stationary Schroedinger equation, taking as arguments the instantaneous configurations of the Universe, $\phi$. From this equation, according to some proposals for quantum gravity, Time should emerge relationally --- for one system wrt another --- or not at all (see e.g \cite{Rovelli_timeless, Kiefer, Rovelli_book}). Another topic which tangles quantum gravity and quantum foundations, and is particularly relevant for decoherence properties  (as in e.g. envariance \cite{Zurek_envariance}),  is that once gravity is included the locality of subsystems and the respective tensor product factorization of subsystems  become much more subtle \cite{Hartle_Giddings, Donnelly_local, Donnelly_2016}. 
 
The avenue of investigation pursued in this paper takes timelessness to heart,  blurring the distinction between unitary evolution and reduction of the wave-function. Unlike previous attempts in which reduction processes are effectively included in the unitary evolution, here both types of processes emerge from a fundamentally timeless description of the entire Universe. It is the fundamental timelessness of the theory which does away with the necessity of a physical wave-function collapse. Moreover, subsystems, including their use in envariance, should emerge dynamically. This emergence is addressed in another paper \cite{Locality_riem} and a precise relation with envariance will be elucidated in a forthcoming paper. 

As I will describe, timelessness here implies that physics is encoded in a static volume-form over the physical configuration space of the entire Universe. All possible field configurations (which respect given relational symmetry principles) will occur, but the volume-form dictates which type of configurations are more represented than others. 
Then, through a fundamental asymmetry of the reduced configuration space, a notion of time can become available. That is because this asymmetry allows for the formation of records in the volume-form. These records provide ordering of configurations, and these orderings  enable time. The emerging picture allows a  reconciliation between a naive timeless Schroedinger interpretation of the wave-function with the one coming from the Conditional Probabilities Interpretation \cite{Page_summary, Dolby, Rovelli_book}. 

 Nevertheless, there is a picture of decoherence that is appropriate to this global setting. Since, the approach should contain a description of the entire Universe, this picture bears strong parallels to formulations of \textit{consistent histories} (see \cite{CH_review} for a review) through path integrals in configuration space. The present work thus inherits certain objections to consistent histories as posed in that setting, e.g.:  Halliwell\rq{}s  quantum zeno paradox for coarse-grainings of paths by regions in configuration space \cite{Halliwell_Zeno}, and the \lq\lq{}set selection problem\rq\rq{} alluded to in the abstract. By incorporating timelessness, configuration space acquires the status of a preferred basis. Within this setting, coarse-grainings of histories can be uniquely given while embodying classical behavior in a more straightforward sense than standard consistent histories. In sum,  I will show that for the purposes of this paper, semi-classical coarse-grainings in configuration space can be constructed which avoid standard criticisms.

Since the present approach also stipulates that the entire space of configurations actually exists, I must also deal with some of the standard objections to the Many Worlds interpretation. In particular, I need to make the meaning of the Born rule explicit. This will be achieved along the lines described in the summary below; all configurations exist, but different regions can have different volumes and thus events can be more or less represented in the space of all configurations.  

\subsection{Summary of the setup.}\label{sec:summary}
The claim being made in this paper is that a timeless perspective on quantum mechanics can potentially resolve the measurement paradox. Time \textit{can} still emerge from a fundamental asymmetry in configuration space \cite{Barbour94_2}.

The fact is that a complete resolution of the measurement problem still evades us. The  obstruction to a realist resolution by the Many-Worlds approach (aided by decoherence), is the lack of \lq\lq{}definite outcomes\rq\rq{}. To put it bluntly, in Many-Worlds, there is no justification for why branches occur, and with a frequency given by the Born rule for probabilities. Many attempts exist which attempt to embody such  frequency in a \lq\lq{}decision-theoretic\rq\rq{}  framework (see \cite{Wallace_book} for a review). However, it seems to me that the arguments in this direction are still unconvincing to a majority of physicists. Moreover, there are still unresolved issues in making Many-World branching compatible with a block-Universe view \cite{Saunders_Time}.

In this paper,  I propose a distinct resolution, realist and not single-world. It consists of stipulating that what truly exists is the set of all configurations $\mathcal{Q}$, endowed with a particular volume-form $\mu:\mathcal{Q}\rightarrow \mathbb{R}$ on top of it: 
$$V(R):=\int_R \mu(q)*1$$ for a given region $R\subset \mathcal{Q}$, and a heuristic Hodge star operator, $*$.\footnote{Here $*$ refers to an arbitrary smooth Riemannian metric on $\mathcal{Q}$. For my purposes here it is enough to leave it at this level of abstraction. For more detail see \cite{QG_deco}.} The volume-form provides a means to count configurations; it is a minimal object required to have a statistic interpretation of physical phenomena.

In this perspective all events occur -- because they are all represented in configuration space -- but they occur with different frequencies. In other words, suppose event $A$ is  equally represented in all the configurations in a region $R_A$ -- meaning that all of these configurations contain some subsystem in a  state which we would identify, up to a certain accuracy, as $A$ -- and event $B$ is likewise represented in all configurations in $R_B$. Then the relative volume (or \lq{}number\rq{}) of the regions is given by $V(R_A)/V(R_B)$.   This provides a simple resolution to the \lq\lq{}definite outcomes\rq\rq{}  problem -- all events occur -- and a straightforward way to introduce probabilities into physics -- these events are represented by different numbers of configurations. 

In the paper \cite{QG_deco}, I also give a particular realization of such a volume-form which recovers the Born rule. It recovers the Born rule for a single, static wave-function, which, as I argue, is given uniquely by simple first principles. For the purposes of this paper, the important principles are:
\begin{enumerate}
\item A preferred element of configuration space exists, $\phi^*$.  This axiom  embodies the required fundamental asymmetry of (reduced) configuration space, and replaces the need for a Past Hypothesis. 
\item Constructing the path integral for a given action functional respecting the instantaneous symmetries, $S(\gamma)$, where $\gamma:\mathbb{R}\rightarrow \mathcal{Q}$, this principle is simply that the volume-form be given by 
\be \mu(q):=F(W(q_o,q))
\ee
where 
\be\label{transition_amplitude} W(q_o, q)=\int \mathcal{D}\gamma\exp{(i S(\gamma))}\ee is the transition amplitude between the initial point of all paths in the path integral --- which can be identified with the  least complex instantaneous state \cite{QG_deco} --- and $q$, 
for some $F:\mathbb{C}\rightarrow \mathbb{R}_+$, which has a factorization property: 
\be\label{equ:factorization}
F(z_1z_2)=F(z_1)F(z_2)
\ee 
This property is required so that amplitude factorization is reflected on the volume form. It embodies a Markovian property of probabilities. 
\end{enumerate}

 One important point of the construction is that  the volume-form can have certain correlations. Such correlations embody the concept of conditional probability, and are the basis of my definition of records. In extreme cases, these correlations correspond to  \lq{}creases\rq{} in (reduced) configuration space: curves where the volume-form concentrates. These creases allow records within configurations to be reliably seen as the outcome of a dynamical evolution. A multitude of records for different subsystems seem to be conversant, and to have jointly evolved from an earlier configuration. For all practical purposes, this is what we interpret as the past history of the Universe. 
 
In this timeless  view, observers are part of the overall configuration. The whole of physics should be encoded in a static probability density over configuration space. I interpret such a density simply as a volume-form -- a way to \lq{}count\rq{} a number of configurations in a given region. This density might reflect a dearth of particular observers; for instance, those that would cohabit a world in which an electron hits a destructive-interference spot in a double-slit experiment.  That is, configurations with multiple records of the same event can  concentrate the volume form on a given region, $A$;  the volume supported in a region with similar records but contained in $\mathcal{Q}-A$ can be tiny.  
 Lastly, the construction of records allows one to establish a notion of conservation of probability.

\section{Path integrals in configuration space}\label{sec:config_space}
\subsection{Configuration space}\label{sec:observers}
\begin{quote}
``Indeed, all measurements of quantum mechanical systems could be made to reduce eventually to position and time measurements (e.g., the position of a needle on a meter or time of flight of a particle). Because of this possibility a theory formulated in terms of position measurements is complete enough to describe all phenomena." - Feynman and Hibbs \cite{Feynman_Hibbs}
\end{quote}
 I would add that even the reading of a clock is abstracted from a position measurement; we often identify it with a given configuration of that common contraption  on the wall (or on our wrists). These facts are argued for extensively in favor of a representation of physics in configuration space in Barbour's `The End of Time' \cite{Julian_End} (although there the arguments are still couched on a Wheeler-DeWitt interpretation of quantum gravity).  
 There are good reasons to assigning preferred status to a position representation of quantum mechanics. Of course, in the end one can translate the formalism to momentum space (e.g. if a Fourier type of transformation is available), but here the primary physical ontological entity is taken to be configuration space, with other aspects being alternative epistemological descriptions only. 

Examples of configuration space can be given by the positions of $n$-particles in a $d$-dimensional space,  $\mathcal{Q}=\{q_1, \cdots q_n\}\simeq \mathbb{R}^{dn}$, or in the field space case, as a space of sections of a given tensor bundle over a finite-dimensional manifold $M$, e.g. $\mathcal{Q}=C^\infty(TM\otimes_i TM\otimes TM^*\otimes_j TM^*)$, for the $(i,j)$ tensors over $M$. In the gravitational case, this is taken as a subspace of such a tensor bundle, the subspace of positive symmetric $(0,2)$-tensors, $\mathcal{Q}=C_+^\infty( TM^*\otimes_S TM^*)$ which forms a \lq\lq{}cone\rq\rq{} inside of the vector space  $C^\infty( TM^*\otimes_S TM^*)$. In the general case, I will denote elements of configuration space simply  by $\phi\in \mathcal{Q}$.

\paragraph*{Observers as submanifolds in configuration space}

 There is a type of  reciprocity between physical space and configuration space which can be described as follows. 
Whereas fixing the entire field configuration (e.g. non-locally on $M$) defines a point in the configuration space $\mathcal{Q}$, fixing only a partial field configuration (e.g. on a subset of $M$) determines an entire submanifold of $\mathcal{Q}$.\footnote{To be more careful, I should only ascribe to them the title of `subsets', not submanifolds. However, under reasonable assumptions, as shown in the accompanying paper \cite{Locality_riem} they indeed form submanifolds.} Such submanifolds are formed by all of the configurations which have that same fixed field, let's say $\phi_{\mbox{\tiny{O}}}$ defined on $O\subset M$,  i.e. those fields $\phi$ which coincide with $\phi_{\mbox{\tiny{O}}}$ on  $O$ but are arbitrary elsewhere: 
$$\mathcal{Q}_{\phi_{\mbox{\tiny{O}}}}:= \{\phi\in \mathcal{Q}~|~\phi_{|\mbox{\tiny{O}}}= \phi_{\mbox{\tiny{O}}}\}
$$
This is easily seen in the finite-dimensional case, in which $\mathcal{Q}=\{q_1, \cdots q_n\}\simeq \mathbb{R}^{3n}$ and the \lq{}region\rq{}  is defined by a subset of particles having a fixed value, e.g. $\{q_i, \cdots q_{i+k}\}=\{q^o_i, \cdots q^o_{i+k}\}$, and is thus isomorphic to an embedding  $\mathbb{R}^{3k-n}\hookrightarrow \mathbb{R}^{3n}$. 
  Alternatively, one could have a subset in a given range, e.g. of $k$ particles in a given \lq\lq{}cube\rq\rq{},  $\{q_i, \cdots q_{i+k}\}\subset [-a,a]^{3k}\subset \mathbb{R}^{3k}$, or more complicated submanifolds, defined by regular values of smooth submersions $f:\mathbb{R}^{3n}\rightarrow \mathbb{R}^{m}$, for $m<3n$; this would be more appropriate for relationally defined regions, where the function $f$ defines the relation between the \lq{}observed particles\rq{}. A simple example would be e.g.: all those configurations for which the relation between the interparticle separation between three of the particles is the same: $f(q_1,\cdots, q_n):=|q_k-q_{k-1}|-|q_{k-1}-q_{k-2}|=0$, which is a regular submersion and thus defines a 3n-1 dimensional surface. More complicated surfaces and, intersection of surfaces, can be obtained by such relational submersions.

Here I thus consider \lq{}instantaneous observers\rq{} to be defined by specific partial field configurations.  There are no subjective overtones  attributed to an observer -- they are merely  partial states of the fields and thus  represented by such submanifolds as briefly described above and at length in the accompanying paper \cite{Locality_riem}. There it is shown how tensor product factorizations can emerge not for the wavefunction, but for transition amplitudes, $$W(\phi_1, \phi_2)\simeq W_O(\phi^O_1, \phi^O_2)W_{O\rq{}}(\phi^{O\rq{}}_1, \phi^{O\rq{}}_2)$$ 
where the manifold decomposes as $M=O\cup O\rq{}$, i.e. if the subsystem of interest is $O$, then $O\rq{}$ would be analogous to \lq{}the environment\rq{}, and the transition amplitude factorization leads to an analogue of a tensor product factorization of the Hilbert space (and is thus amenable to einselection arguments \cite{Zurek_envariance}, which will be explored elsewhere). 

    One could form a submanifold in $\mathcal{Q}$ by restricting to all those points where an observer\rq{}s brain state is correlated with some other part of the field. Of course, there are many parts of configuration space where no such thing as \lq{}an observer\rq{} will exist. 



\subsection{The timeless  transition amplitude in quantum mechanics}\label{sec:timeless_ta}
According to Wheeler, \lq\lq{}the past has no existence except as it is recorded in the present\rq\rq{}. {Even within standard approaches, it seems plausible to idealize each observation as occurring within a single spatial hypersurface and having a certain spatial extent \cite{Page_summary}.  } Very generally, following Page \cite{Page_summary}, if the quantum state of the Universe on a spatial hypersurface is given by a density matrix $\rho$, the conditional probability of $A$, given a testable condition $B$, is:
$$\mu(A|B)=\frac{\mbox{Tr}(P_AP_B\rho P_B)}{\mbox{Tr}(P_B\rho P_B)}$$
where $P_A=P_A^\dagger=P_A^2$ is a corresponding projection operator  (mutatis mutandi for $B$). 

As one does not have access to coordinate time, in the presence of a Hamiltonian one should average over this inaccessible variable to get a density matrix for the physical variables. I.e. one needs to use the projector
\be\label{equ:projector}
\hat P:=\lim_{\tau\rightarrow\infty}\frac{1}{2\tau}\int_{-\tau}^\tau d\tau\, e^ {-i\tau \hat H}
\ee
on the density matrix and on the other operators, where $\hat H$ is the canonically quantized (with Weyl ordering)  Hamiltonian. For instance, $\hat A\mapsto \hat A_{\mbox{\tiny phys}}=\hat P \hat A\hat P$. 

 Here $\tau$ is the flow parameter of the Hamiltonian, and should not strictly speaking be thought of as \lq\lq{}time\rq\rq{}. In the path integral representation, paths in configuration space will be allowed to be non-monotonic in this parameter.  Note that since one integrates over all $\tau$, the projector is parametrization independent. For a careful analysis of the emerging quantum mechanical description of \lq\lq{}timeless\rq\rq{} physical systems, see \cite{Dolby}. There, it is shown explicitly how to make sense of the probability mentioned in the introduction, of $\mu(A_2~~ \mbox{when}~B_2~| ~A_1 ~~\mbox{when}~B_1)$, in a timeless fashion (including a rebuke to Kuchar\rq{}s objection \cite{Page_summary}) and how to recover the standard quantum mechanics transition amplitude in the presence of a good clock subsystem. Such relational timeless interpretations are also related to Rovelli\rq{}s ideas (see \cite{Rovelli_timeless} and \cite{Rovelli_book}). 

Here, I am interested in a timeless transition amplitude for path integrals  in configuration space, which is more directly applicable to the axioms and motivations of this work, as presented in section \ref{sec:summary}. The work of Page and others can be cleanly transposed to the path integral setting, at least in the  particle mechanics case \cite{Chiou}.  The role of records is to unify a conditional probabilities interpretation with a naive timeless Schroedinger interpretation, through a fundamental asymmetry of the volume form in configuration space, or alternatively, through special boundary conditions for the wavefunction.

\paragraph*{ Timeless transition amplitude for path integrals  in configuration space}
Let $\mathcal{Q}$ be the configuration space of  a finite-dimensional system,    coordinitized by $q^a$, for $a=1,\cdots, n$. No coordinate, or function of coordinates, need single itself out as a reference parameter of curves in $\mathcal{Q}$, curves which one might associate with the evolution of the system. The systems we are considering are not necessarily `deparametrizable' -- they do not necessarily possess a suitable notion of time variable applicable everywhere in configuration space.  

Defining the fundamental transition amplitudes between configuration eigenstates:\footnote{As much as possible, I want to avoid technicalities which won't be required here. Having said this, formally one would have had to define the so-called kinematical Hilbert space $\mathcal{K}$ for the quantum states over $\mathcal{Q}$ by using a Gelfand triple over $\mathcal{Q}$  with measure $d^dq^a=dq^1\cdots dq^d$, i.e.  $\mathcal{S}\subset\mathcal{K}\subset\mathcal{S}'$. This is not necessary in my case, because we will not require a Hilbert space.}
\be
\label{equ:transition}
W(q_1,q_2):=\langle q_1|\hat P|q_2\rangle
\ee 
Without absolute time, one must employ new tools in seeking to show the equivalence. For instance, a parametrized curve $\bar\gamma:[0,1]\rightarrow \Omega$ need not be injective on its image; it may go back and forth in whatever partitioning of paths $\tau$ one chooses.  
 Chiou uses a Riemann-Stieltjes integral as opposed to a Riemann one in order to make sense of the limiting procedure to infinite sub-divisions of the parametrization \cite{Chiou}. 
 
 To be more specific, for a given $\tau$, one introduces a sequence, $\tau_i$, $i=1,\cdots N$, such that $\tau_0=0, \tau_N=\tau$. The mesh of the sequence is defined as $\max_{i}|\Delta\tau_i|$, where $\Delta\tau_i:=\tau_i-\tau_{i-1}$, and for building the path integral one takes limits where the mesh is vanishingly small. This will depend on a choice of mesh, so, in the end, we implement a gauge-averaging procedure \cite{Marolf} (which can be done when the group is just $\mathbb{R}$, as is our case, even in field theory \cite{QG_deco}).

 Using the completeness relations of both the momenta and the position eigenstates (i.e. decomposition of the identities), we write
 $$ 
 \langle q_1^a|e^{i\Delta\tau \hat H}|q_2^a\rangle=\left(\prod_{n=1}^N\int d^dq_n^a\right) \langle q_N^a|e^{i\Delta\tau \hat H}|q_{N-1}^a\rangle \langle q_{N-1}^a|e^{i\Delta\tau \hat H}|q_{N-2}^a\rangle\cdots \langle q_1^a|e^{i\Delta\tau \hat H}|q_{0}^a\rangle
 $$
Note that we have $N$ copies of configuration space integrated over, which are not at this point associated to different times.  For small mesh, we can expand 
 $$\langle q_1^a|e^{i\Delta\tau \hat H}|q_2^a\rangle\approx \langle q_1^a|1+{i\Delta\tau \hat H}|q_2^a\rangle$$
  and with the additional fact that, with Weyl ordering, the following relation holds: 
 $$\langle q_i^ a| { \hat H (\hat q, \hat p)}|q_{j}^a\rangle=\int \frac{d^dp_a}{(2\pi\hbar)^d}\exp\left(\frac{i}{\hbar}p_a(q_i^a-q_j^a)\right)H\left(\frac{q_i^a+q_j^a}{2},p_a\right)
 $$ we find in the limit where the mesh goes to zero (using a Riemann-Stiltjes integral): 
 \be\label{equ:PI_step}
  \langle q_1^a|e^{i\tau \hat H}|q_2^a\rangle=\lim_{N\rightarrow \infty}\left(\prod_{i=1}^N\int d^dq_i^a\right)\left(\prod_{i=1}^N \int \frac{d^dp_{ia}}{(2\pi\hbar)^d}\right)\exp\left(\sum_{i=1}^{N-1}\frac{i}{\hbar}p_{ia}(\Delta q^a_i)\right) \exp{\left(i\sum_{i=0}^{N-1}\Delta \tau_{n+1}H\left( \bar q_i^ a,p_{ia}\right)\right)}
 \ee
 where $\bar q_i^ a:=\frac{q_i^a+q_j^a}{2}$ and note that $i$ is not a spacetime index. Now, we have a sequence of phase spaces being integrated over: $\prod_iT^*\mathcal{Q}_i$. In the limit, we take arbitrary sequences $\{(p_i, q_i)\}_{i=1,\cdots N}$, through a relabeling $i\rightarrow \tau_i$,  as a continuous curve $(q^a(\tau), p_a(\tau))$ (possibly with self-intersections) projected to a single phase space $T^*\mathcal{Q}$, and with fixed projected endpoints on $\mathcal{Q}$ fixed to $q_1^a$ and $q_2^a$. 
 
 In more standard notation, by e.g. identifying 
 $$ \sum_{i=1}^{N-1}p_{ia}\Delta q_{i^a}\rightarrow \int_{\bar\gamma} p_a dq^a\equiv \int_{\bar\gamma}p_a \frac{dq^a}{d\tau\rq{}}d\tau\rq{}\qquad\mbox{and}\qquad  \sum_{i=0}^{N-1}\Delta \tau_{n+1}H\left( \bar q_i^ a,p_{ia}\right)\rightarrow \int_{\bar\gamma} H(q^a(\tau\rq{}), p_a(\tau\rq{}))d\tau\rq{}
 $$ for each choice of the sequence $p_i, q_i$ resulting in $\bar\gamma(\tau\rq{})=(q^a(\tau\rq{}), p_a(\tau\rq{}))$, such that in this parametrization $q^a(\tau\rq{}=0)=q^a_1$ and $q^a(\tau\rq{}=\tau)=q_2$, then \eqref{equ:PI_step} becomes:
  \be\label{intermediary} \langle q_1^a|e^{i\tau \hat H}|q_2^a\rangle=\int\mathcal{D}q^a \mathcal{D}p_a \exp\left(\frac{i}{\hbar}\int p_a(d q^a)\right) \exp{\left(i\int_{\bar\gamma} H\left( \bar q^ a(\tau\rq{}),p_a(\tau\rq{})\right)d\tau\rq{}\right)}\ee
 Notice that the only place where the parametrization appears is inside the Hamiltonian, the first term of \eqref{intermediary} is reparametrization invariant. However, as noted, we have chosen a particular mesh, and now we must use group averaging \cite{Marolf} over $\mathbb{R}$ to project onto the gauge-invariant states. Thus we parametrize the mesh with lapses, writing for each interval $\Delta \tau_n =\hbar^{-1}N_n\Delta \tau_n\rq{}$.

   Doing the averaging in this parameter, and integrating also over $\tau$,  yields (reverting back to the summation, as opposed to the integration, of the Hamiltonian term): 
 \be\label{step_x} W(q_1,q_2)=\int\mathcal{D}q^a \mathcal{D}p_a  \mathcal{D}N \exp\left(\frac{i}{\hbar}p_a(d q^a)-\frac{i}{\hbar}\sum_{n=0}^{N-1}\Delta \tau\rq{}_{n+1}N_{n+1} H\left( \bar q^ a(\tau\rq{}),p_a(\tau\rq{})\right)\right)
 \ee where 
 $$\prod_{n=0}^{N-1}\int_{-\infty}^\infty dN_{n+1}\rightarrow \int \mathcal{D}N
 $$ Finally, since in the continuous limit the sum converges to the integral: 
 $$ \sum_{n=0}^{N-1}\Delta \tau\rq{}_{n+1}N_{n+1} H\left( \bar q^ a(\tau\rq{}),p_a(\tau\rq{})\right)\rightarrow \int_{\bar\gamma}N(\tau\rq{})H\left( q^ a(\tau\rq{}),p_a(\tau\rq{})\right)d\tau\rq{}
 $$
 we obtain, performing the lapse integral in \eqref{step_x}: 
 \be\label{equ:discrete_timeless_PI} W(q_1,q_2)=\int\mathcal{D}q^a \mathcal{D}p_a \exp\left(\frac{i}{\hbar}p_a(d q^a)\right)\delta(H)\ee
The remarkable property of this formula is that only the \textit{curve} in phase space (i.e .the unparametrized path) factors in. The $\delta(H)$ term is what in this case we associate with the \lq\lq{}gauge-symmetry\rq\rq{} of reparametrization invariance.

If the Hamiltonian is quadratic in the momenta --- as in standard dynamical systems with time --- one can integrate them out obtaining a Lagrangian path integral on configuration space \cite{Chiou}. Once again, it is only the curves in configuration space that are relevant, not their parametrization. 

   If configuration space can be decomposed wrt a time-like variable, $q=(t,\bar q)$, such that  the total Hamiltonian can be written in a $t$-dependent way as $\hat{H}=\hat{p}_t+\hat{H}_o(\bar q, {p}_{\bar q}, t)$, then two alternatives arise: i) the commutator at two different times is zero, $[H_o(t_1), H_o(t_2)]=0$. In this case the timeless path integral amplitude kernel is equal to the standard non-relativistic one,  up to an overall constant:\footnote{In \cite{Briggs_Rost},  the conditions under which Briggs and Rost derive the time dependent Schroedinger equation from the time-independent one, corresponds to the system being deparametrizable. I.e. one can isolate degrees of freedom (the environment, in Briggs and Rosen) which are heavy enough to not suffer back reaction. It is a stronger condition, in that there $H_o$ is supposed independent of $t$.}
\be\label{equ:deparametrizable}
W(q_1,q_2)=G((t_1,\bar q_1),(t_2,\bar q_2))=\langle\bar q_1|e^{-i\hat{H}_o(t_2-t_1)}|\bar q_2\rangle
\ee
As remarked on by Chiou, this means that although the timeless system is kinematically different than the non-relativistic one, it is still dynamically equivalent.  Note moreover, an important difference: in the lhs of \eqref{equ:deparametrizable}, time is just part of the configuration, it has no role as a \lq\lq{}driver of change\rq\rq{} \cite{CFP_essay}. In the remaining parcels  of the equation, it is interpreted as an absolute time.  

The second alternative, ii) is that $[H_o(t_1), H_o(t_2)]\neq 0$. In that case, the system is both kinematically and dynamically distinct from the non-relativistic one. Here, paths that go back and forth in the parametrization $t$ of the curves become relevant, thus distinguishing the timeless path integral from the standard path integral, with time-ordered paths. Nonetheless, in the semi-classical approach, the distinction becomes suppressed, because extremal paths are single-valued in $t$ if $\hat{H}=\hat{p}_t+\hat{H}_o(\bar q, {p}_{\bar q}, t)$ \cite{Chiou}.


 \subsection{The semi-classical transition amplitude}\label{sec:semi_classical}
In the context of path integrals in configuration space, I will be in the \emph{semi-classical} (or WKB, or saddle point), approximation. 
 
 We are in the context of an oscillatory path integral in configuration space, \eqref{transition_amplitude}, for (locally) extremal paths parametrized by the set $I\subset \mathbb N$,  $\{\gamma_{\mbox{\tiny{cl}}}^\alpha\}_{\alpha\in I}$, between an initial and a final field configuration $\phi_i,\, \phi_f$. Denoting the on-shell action for these paths as $S_{\gamma_{\mbox{\tiny{cl}}}}$, the expansion, accurate for $1<< S_{\gamma_{\mbox{\tiny{cl}}}}/\hbar$, can be written (see \cite{Chiou} for a proof in finite dimensions):  
\be\label{equ:semi_classical_exp} W_{{\mbox{\tiny{cl}}}}({\phi_i}, \phi_f)= A \sum_{\alpha\in I}(\Delta_{\gamma^\alpha_{\mbox{\tiny{cl}}}})^{1/2}\exp{\left(i S({\gamma^\alpha_{\mbox{\tiny{cl}}}})/\hbar\right)}
 \ee 
 where $A$ is a normalization factor (independent of the initial and final configurations). {In a finite-dimensional dynamical system of dimension $d$, this takes the form of the phase space volume occupied by a quantum state: $A=(2\pi\hbar)^{-d}$. } The Van Vleck determinant is defined as
 \be\label{equ:path_Van_vleck}
 \Delta_{\gamma_{\mbox{\tiny{cl}}}}^\alpha:=\det\left(-\frac{\delta^2 S({\gamma^\alpha_{\mbox{\tiny{cl}}}})}{\delta {\phi_i}\delta\phi_f}\right)= \det\left(-\frac{\delta\pi^\gamma_f[{\phi_i}, \phi_f]}{\delta {\phi_i}(y)}\right)
 \ee
where  the on-shell momenta is defined as 
\be\label{mom_VV}\pi^\gamma_f[{\phi_i}]:= \frac{\delta S_{\gamma^\alpha_{\mbox{\tiny{cl}}}}[{\phi_i},\phi_f]}{\delta \phi_f}\ee
where we here write the action as a functional of its initial and final points along ${\gamma_{\mbox{\tiny{cl}}}}$. 

If we use the classical trajectories to transport the points in an infinitesimal volume around $\phi_i$, defining an infinitesimal volume around $\phi_f$, the Van-Vleck determinant can be interpreted as a ratio of these volumes, measured by the relative density of extremal curves around $\phi_i$ and $\phi_f$, 
\be\label{equ:Van_vleck_volume}
\Delta_{\gamma_{\mbox{\tiny{cl}}}}^\alpha=\frac{\rho_\alpha(\phi_f)}{\rho (\phi_i)}
\ee
In an analogy with Riemannian geometry, the Van Vleck determinant would be related to the integral of the expansion scalar along a geodesic congruence. 

 Assuming that there exists at least one locally extremal (classical) path between the two configurations: 
\be\label{equ:semi_field_interference}
|W_{{\mbox{\tiny{cl}}}}(\phi_i,\phi_f)|^2=A\left(\sum_{\alpha\in I}\Delta_{\gamma^\alpha_{\mbox{\tiny{cl}}}}+2\sum_{\alpha\neq\alpha\rq{} }|\Delta_{\gamma^\alpha_{\mbox{\tiny{cl}}}}\Delta_{\gamma^{\alpha\rq{}}_{\mbox{\tiny{cl}}}}|^{1/2}\cos{\left(\frac{ S_{\gamma^\alpha_{\mbox{\tiny{cl}}}}-S_{\gamma^{\alpha\rq{}}_{\mbox{\tiny{cl}}}}}{\hbar}\right)}\right)
\ee 
where we have omitted the dependence of the Van-Vleck determinant on the configurations.  Here interference terms can be clearly identified: they are the cosine terms of the difference between the on-shell action of different classical paths and weighed by the Van-Vleck determinants.

Note that in this formulation, the very construction of the amplitude square already bears the effect of interference, even on a timeless setting. Now I will quickly show that the Born rule uniquely emerges as the volume form defined by the amplitude and the factorization property \eqref{equ:factorization}.  
 
  \subsection{The Born rule}\label{sec:Born}

 In the semi-classical regime, we can associate the Born rule with a relative `density of observers' if we interpret  the likelihood of finding oneself e.g. in a region around configuration $\phi_1$ relative to configuration $\phi_{2}$ as the relative volume of these two regions.

 In the usual decoherence based description of probabilities, a density matrix for which environment degrees of freedom are traced out is diagonalized around pointer states (the action of decoherence can also be used to select the pointer states themselves).  One then obtains probabilities from the diagonal coefficients of such a reduced density matrix. 
 In the present case, to get to an objective meaning of probabilities, I will use a form of diagonalization appropriate to our language. Namely, I will study the case when there  is no significant interference between different elements of a particular semi-classical coarse-graining. 
 
 In this regime, we can use \eqref{equ:semi_field_interference}, and compare our density function $\mu(\phi)$ to  densities propagated by classical dynamics in configuration space. 
From  \eqref{equ:Van_vleck_volume}, 
\be\rho(\phi_f)=\rho(\phi_i)\Delta_{\gamma_\alpha}
\ee
 Where $\gamma_\alpha$ is the unique extremal path  connecting $\phi_i$ and $\phi_f$. In the general case, one has more than one extremal (or only piece-wise extremal)  path between $\phi_i$ and the final configuration $\phi_f$, and thus we cannot take $\Delta$ to define the densities, since it is path dependent, i.e. it only gives  e.g. $\rho_{\alpha}(\phi_f)$, not  $\rho(\phi_f)$.

 However, for the semi-classical kernel, from the no-interference terms in \eqref{equ:semi_field_interference}, we should also have that $\Delta_{\gamma_\alpha} \propto |W_{\mbox{\tiny cl}}(\phi_i,\phi_f)|^2$, and thus 
 \be \label{equ:Born_semi} \rho(\phi_f)=\rho(\phi_i)  |W_{\mbox{\tiny cl}}(\phi_i,\phi_f)|^2
\ee
This already restricts the semi-classical, no interference limit of the probability density to be given by the Born rule. {Since we will have a unique choice of $\phi_i$, given by the preferred \lq{}vacuum\rq{}, or in-state $\phi^*$, we can also absorb $\rho(\phi^*)$ into the normalization $A$. }

As I mentioned, for records to have the correct factorization property, one only needs to assume that the positive function $F:\mathbb{C}\rightarrow \mathbb{R}^+$,  given in axiom 3 of \ref{sec:summary}, has property \eqref{equ:factorization}, i.e.; that $F(z_1z_2)=F(z_1)F(z_2)$.  However, it can be shown that this factorization property implies that $F(z)=z^\alpha \bar z^\beta$, for arbitrary numbers $\alpha, \beta$. {The proof requires only that one take the derivative wrt to $z_1$ and set $z_1=1, z_2=z$, obtaining a differential equation which only has a homogeneous solution, and then do the same with the conjugated variables.  Since $F$ needs to be positive, $\alpha=\beta$. Then, together with the semi-classical limit above, we necessarily get $\alpha=1$, i.e. the Born rule.  In other words, for my definition of records to have a conditional probability property and the required semi-classical interpretation in terms of propagated volumes,   the Born rule emerges uniquely. 
 
To reiterate, the full  volume form  related to the \emph{static wave-function over configuration space} $\psi(\phi):= W(\phi^*,\phi)$, is\footnote{Alternatively, we could have defined a \lq\lq{}vacuum state\rq\rq{} $|\phi^*\rangle$ existing over a trivial (one complex dimension) Hilbert space $H_{\phi^*}$ with the usual complex space inner product,  over $\phi^*$,  and  then taking $\hat W(\phi^*,\phi):H_{\phi^*}\rightarrow H_{\phi}$ as an operator such that $|\psi(\phi)\rangle:= \hat W(\phi^*,\phi)|\phi^*\rangle$.}
  \be\label{Born} \mu(\phi)=|W(\phi^*,\phi)|^2\ee

\section{Decoherence and coarse-grainings}\label{sec:decoherence}
Here I have given a definition of a single probability density over field space, without a parameter time and consequently without the non-unitary intervention of \lq\lq{}measurements\rq\rq{}. Nonetheless, decoherence -- our best attempt at explaining within standard quantum mechanics how unitary processes can blend into non-unitary ones --  still plays an important role, which I now explain.

In the present formulation, which is in its essence a form of Many-Worlds, decoherence properties are recovered dynamically  as lack of interference between different families of paths, precisely as it is done in the consistent histories setting (see \cite{CH_review, Hartle_GM} and references within). The analogy requires a particular \lq\lq{}framework\rq\rq{} in the language of consistent histories; namely the \lq{}framework\rq{} of paths in configuration space. The usual issue with CH is that one could choose different, complete sets of non-commuting coarse-grained histories to represent a given process. The limitation that a representation be given as paths in configuration space severely limits this problematic degeneracy. Moreover, different types of measurement apparata belong to different regions of configuration space, and thus will represent fundamentally different processes (an example of \lq\lq{}contextuality\rq\rq{}). I will discuss this further in the conclusions. 

  In the present instance, the decoherence functional of consistent histories, for a given choice of sets of coarse-grained paths $C_{\alpha}=\{\gamma_{I_\alpha}\}$, ($I_\alpha$ the parametrization of paths in $C_\alpha$) between $\phi^*$ and $\phi$,  is given by: 
\be\label{equ:new_deco_path} D(\alpha_1, \alpha_2)=\int_{\gamma_1\in C_{\alpha_1},\gamma_2\in C_{\alpha_2}}  \mathcal{D}\gamma_1(\tau)\mathcal{D}\gamma_2(\tau)e^{i(S[\gamma_1(\tau)]-S[\gamma_2(\tau)])/\hbar}\ee  The condition for a coarse-graining to be deemed \emph{consistent}, i.e. for it to yield classical probabilities, is for the decoherence functional \eqref{equ:new_deco_path} to vanish, signifying decoherence between the alternative coarse-grained sets.

  The main benefit of a treatment through consistent histories, which I can recycle here, is that it is applicable without the need for a separation between system and environment. However, I need to address one of the outstanding questions of consistent histories: what are the coarse-grained sets of paths that form them? In clarifying this issue, as the formulation is done entirely within configuration space, there is one problem,  raised by Halliwell \cite{Halliwell_Zeno}: do the most natural coarse-grainings in configuration space suffer from the quantum Zeno effect? I will here sketch a solution to this issue.

Moreover, I should stress two things here: firstly,  unlike consistent histories, the (coarse-grained) histories themselves are not observable here, even if decohered (a view supported by \cite{Page_summary}). The observables are configurations and possibly records therein. Secondly, the brunt of the work for decoherence is done merely by having paths for which the action is much greater than $\hbar$, according to the Imaging Theorem \cite{Imaging_theo}. Distinctly from environmental
decohering interaction, this can provide the transition necessary for an observer to interpret
perceived quantum dynamics as classical.


\subsection{Coarse-grained histories}

Translated to this context, fine-grained histories are just piecewise smooth curves in configuration space $\gamma:I\rightarrow \mathcal{Q}$.  Coarse grained histories are partitions of all the fine grained histories from an initial point of $\mathcal{Q}$ to a final one into bundles of paths, with certain characteristics.

\subsubsection{Extremal coarse-grainings}\label{sec:PEC}

Given a one-parameter family of paths between $\phi_i, \phi_f$, i.e.  $\gamma(u)\in \Gamma(\phi_i, \phi_f)$, such that  $\gamma(0)=\gamma_{\mbox{\tiny{cl}}}$, $\frac{d}{du}_{|u=0}\gamma(u,t)=X(\tau)$ (or $\gamma\rq{}(0)=X$) and $$\frac{d}{du}_{|u=0}S[\gamma(u)]=\delta S\cdot X=0~~~(\gamma_{\mbox{\tiny{cl}}}~\mbox{is extremal)}$$
  It was shown in \cite{Cecille} that to any order in $\hbar$, one can  replace the path integration over all paths between $\phi_i$ and $\phi_f$ by an integration over $X\in \mathbb{X}$, the space of deviation vector fields over $\gamma_{\mbox{\tiny{cl}}}\in \Gamma$, where  $\gamma_{\mbox{\tiny{cl}}}\in \Gamma$ can be taken to be arc-length parametrized (in terms of the Jacobi distance). 
  
  Let us for definiteness take the spatial gravitational configuration space, the space of positive sections of the symmetrized tensor bundle $C^\infty_+(TM^*\otimes_S TM^*)$. This is modeled over a  Banach topological vector space, which is metrizable, i.e. there are many notions of distance that we could implement in $\mathcal{Q}$. Suppose we endow it with some Riemannian metric. Now we define
  \paragraph*{Tubular bundles}
  A \emph{tubular neighborhood} of  a given path $\gamma$ in $\mathcal{Q}$ is roughly a small enough tube around $\gamma$ such that the tube doesn't self-intersect. More precisely, a tubular neighborhood of a submanifold $L$ embedded in a Riemannian manifold $N$ is a diffeomorphism between the normal bundle of  $L$  and an open set of $N$, for which the zero section reduces to the identity on $L$. One first defines the exponential map (that the exponential map is still well defined in the infinite-dimensional setting is shown in \cite{Ebin}),  $\mbox{Exp}:T\mathcal{Q}\rightarrow \mathcal{Q}$. Restricting the exponential map  to the normal bundle of $\gamma$ (i.e. to act only on vectors orthogonal to $\gamma'$),   $\mbox{Exp}:(T\gamma)^\perp\subset\mathcal{Q}\rightarrow \mathcal{Q}$, since $\gamma$ is compact, one can always find a maximum radius $\rho_{\mbox{\tiny max}}$ such that there are no self-intersections of the tubular neighborhood, i.e. 
 $\mbox{Exp}$ is a diffeomorphism between $(T\gamma)^\perp_{\rho_{\mbox{\tiny max}}}$ (normal vectors with maximal length $\rho_{\mbox{\tiny max}}$) and its image in $\cal M$. 
  
  \paragraph*{Defining the minimum  radius of the bundle}

  Using the configuration space metric, we have the orthogonal plane  $P:=(T\gamma_{\mbox{\tiny{cl}}})^\perp$. A basis for all the $X\in \mathbb{X}$ can be formed by all the vectors in  $P$. 
  
  Thus, for each unit normal  deviation vector field $X\in P$ from $\gamma_{\mbox{\tiny{cl}}}$,  we can form the one-parameter family of paths $\gamma_X(u)$ given by 
  \be\label{deviation}\gamma_X(u(\tau),\tau):=\mbox{Exp}_{\gamma_{\mbox{\tiny{cl}}}(\tau)}(u(\tau)X(\tau))\ee
where $u(\tau)\in \mathbb{R}_+$ and  $X(\tau)\in P_{\gamma_{\mbox{\tiny{cl}}}(\tau)}$. These paths should not themselves be extremal (unless $u(\tau)=0, \forall \tau$).   We define the set of coarse-grained paths seeded by the family of extremal paths $\gamma^\alpha_{\mbox{\tiny{cl}}}$, with radius $\rho$ as:
\be\label{coarse_graining}
C_\alpha^\rho=\{\gamma^\alpha_X(u(\tau), \tau)\,~|~\, u(\tau)< \rho\}
\ee
where now for each $\alpha$, $X(\tau)\in P_{\gamma^\alpha_{\mbox{\tiny{cl}}}(\tau)}$. This is a particular set of paths defined by an independent open set at each arc-length parameter time $\tau$; it is not defined by \lq\lq{}all those that don\rq{}t enter a given region in configuration space for all times\rq\rq{}.  It should be noted that paths that lie in the complement $(C_\alpha^\rho)^C=\Gamma(\phi_i, \phi_f)-(C_\alpha^\rho)$ do not require reflecting boundary conditions on the given region of configuration space (the image of the paths inside $\mathcal{Q}$); paths that intersect the region are included in $(C_\alpha^\rho)^C$. It doesn\rq{}t constitute a projection operator onto a single region $R$ in configuration space, which is why it avoids Zeno\rq{}s paradox. 

By the results of \cite{Cecille}, given the  extremal paths $\gamma^\alpha_{\mbox{\tiny{cl}}}$ between $\phi_i,\phi_f$, there exists  small enough radii $\rho^\alpha$ such that
\be\label{rho}
W_{\alpha}(\phi_i,\phi_f)\approx \sum_\alpha\Delta_{\gamma^\alpha_{\mbox{\tiny{cl}}}}\exp{\left[iS[\gamma^\alpha_{\mbox{\tiny{cl}}}]/\hbar\right]}\approx W(\phi_i,\phi_f)
\ee where $W_{\alpha}(\phi_i,\phi_f)$  is a sum only over the sets $C_\alpha^\rho$, and
where the approximation works up to orders of $\hbar^2$. 

I would like to define the maximum radius $\rho$ as that for which points outside the tubular bundle become \textit{semi-classically} distinguishable from the points in the seed, $\gamma^\alpha_{\mbox{\tiny{cl}}}$. I.e. those for which $S/\hbar\sim 1$,  
\be
\label{rhomaxbar}
\bar\rho^\alpha_{\mbox{\tiny max}}=\sup_{\rho\in \mathbb{R}^+}\left\{S_{\mbox{\tiny{cl}}}(\gamma^\alpha_{\mbox{\tiny{cl}}}(\tau),\gamma^\alpha_X(\rho, t))\leq \hbar, ~~\forall t \in [0,1]\right\}
\ee
 If the action is of Jacobi type, i.e. if we can use the action to define the geometry,  $S(\gamma)$ is a length functional for some metric in configuration space (see appendix \ref{app:Jacobi_metric}). It thus defines extremal paths as geodesics, and this radius becomes merely $\bar\rho_{\mbox{\tiny max}}^\alpha=\hbar$.

  Now, of course,  there might be intersection between the different sets of coarse-grainings, with paths belonging to more than one $C_\alpha$.  If  this happens at $\rho_{\mbox{\tiny intersec}}$, then we define 
  \be
\label{rhomax}
\rho^\alpha_{\mbox{\tiny max}}=\max\{\rho_{\mbox{\tiny intersec}}, \bar\rho^\alpha_{\mbox{\tiny max}}\}\ee
so that for this radius there are no paths belonging to two elements of the coarse-graining.  Now I define
\begin{defi}[Extremal coarse-grainings (ECs)]\label{def:ECS}  
   An  extremal coarse-graining for the paths in $\Gamma({\phi^*},\phi)$  is a coarse-graining $\{C^{\rho}_\alpha, \alpha\in I\}$ seeded in the extremal paths $\{\gamma^{\mbox{\tiny{cl}}}_\alpha, \alpha\in I\}$ between ${\phi^*}$ and $\phi$. Where each $C^{\rho}_\alpha$ is given  by \eqref{coarse_graining}, with elements $\gamma_X(u)$ given by \eqref{deviation} and  radii given by $\rho_{\mbox{\tiny max}}$ in  \eqref{rhomax}. 
\end{defi} 
 
 Notice as well that this is a constructive definition of the  coarse-graining. While indeed it is true that they never leave a region around the classical path,  it is not necessarily true that they span all of the paths in a given region (and they are not defined to be the subset of all paths that enter or not a given region). Indeed,  Exp may not even be a  diffeomorphism onto the neighborhood given by $\rho_{\mbox{\tiny max}}$. This is contrary to the standard example (and the target of Halliwell\rq{}s criticism), where paths are \textit{defined} to be those that cross a given region (and thus prompt a \lq\lq{}watched kettle\rq\rq{}  paradox).  Also note that a coarse-graining has nothing to do with the system \lq\lq{}being observed\rq\rq{} (or not) in this region.  Accordingly such coarse-graining has its limitations in capturing the full properties of the path integral. Nonetheless, they are more than sufficient for capturing all semi-classical phenomena.

  \paragraph*{The piece-wise context} 
   As discussed in appendix  \ref{app:dewitt}, unlike the case for complete Riemannian manifolds -- where two given points can always be connected by a geodesic -- it is not always the case that two configurations can be connected by a path that extremizes the action.  Any non-smoothness, for example a non-smooth potential in particle dynamics, can imply two given points are not connected by a smooth extremal path. However they will be connected by extremal paths with `corners', as shown by Marsden \cite{Marsden}.   
  If we want to extend the notion of coarse-grained histories that are ``as classical as possible", we would then adapt coarse-grained histories to be piecewise centered on smooth extremal paths as well. 
   
  Incorporating this definition into the piece-wise extremal context prompts me to formulate an alternative approach to singling out a preferred  coarse-graining which reduces to an EC in the simplest case, but, if conjecture \ref{conjecture} from appendix \ref{app:PEC} holds, should be more generally available.   These are the minimal piecewise extremal coarse-grainings (minimal PECs), whose construction I sketch in the appendix, since the main ideas are already contained in the ECs and we will not require it explicitly here. 
 
\subsubsection{Physical significance of EC\rq{}s and decoherence}\label{subsec:EC_deco}
 Assuming that one can use the Jacobi (or action) metric (see appendix \ref{app:Jacobi_metric}) for defining the bundles, the radius has a very relevant  physical content. As explained in the appendix, lengths of paths according to the Jacobi metric are equivalent to the on-shell action.  Furthermore, for generic  systems  the semi-classical amplitude kernel \emph{decreases} with this distance, because the Van-Vleck determinant decreases with the on-shell action for chaotic systems.  The more sensitive the final configuration is wrt the initial momentum, the more diluted the trajectories in the bundle around the classical path will become  and the smaller the weight from this trajectory in the transition amplitude.

In other words, for generic chaotic systems,  the farther in configuration space two  points are, $d(\phi_1,\phi_2)>d(\phi_1,\phi_3)$ (according to geodesic distance) the more sensitive the final configuration is to the initial conditions, and the smaller the Van Vleck determinant
$\Delta_{\gamma_{\mbox{\tiny{cl}}}}({\phi_1},\phi_2)< \Delta_{\gamma_{\mbox{\tiny{cl}}}}({\phi_1},\phi_3)$. In such cases, the amplitude kernel  decreases with the action distance $\rho$ from the seed of the coarse-graining, and the radius has a natural physical meaning.

\paragraph*{Spatial radius vs configuration space radius}

 Using  extremal coarse-grainings,  one should regard the final point $\phi$ as centered on a ball of radius $\rho_{\mbox{\tiny min}}$, within which the ECs cannot distinguish configurations with a good degree of certainty. 
 
 For illustrative purposes, since the semi-classical approximation (for $S_{\gamma_{\mbox{\tiny cl}}}/\hbar <1$) is good up to orders of $\hbar$, let us assume that for a given system with arc-length parametrization, $\order{\rho_{\mbox{\tiny max}}}\sim \order{\hbar}$.\footnote{Note that in this case the action has dimensions of configuration space length, and thus $\rho$ has the right dimensions, and also that the path parameter along each geodesic becomes proportional to the configuration space length.  }
Now we can make an important distinction  regarding distance with respect to the Jacobi metric, used in definition \ref{def:ECS}, and distance in physical space, measured for instance wrt a Riemannian metric $g_{ab}$.

We take the simplest possible example for the Jacobi metric $h_{ab}$, given in appendix \ref{app:Jacobi_metric}, in the case of a  particle of energy $E$ moving on $M$ with background metric $g_{ab}$ and no potential. The Jacobi metric is $h_{ab}=2Eg_{ab}$. Thus, for a radius of $\rho_{\mbox{\tiny max}}\sim \hbar$, and thus $S_{\gamma_{\mbox{\tiny cl}}}/\hbar <1$, the physical radius $\ell$ around the extremal trajectories must obey 
 \be\label{equ:physical_radius}2E \ell\sim \rho_{\mbox{\tiny max}}\sim \hbar \ee 
which means that with low energy/mass wrt $\hbar$ one can have a pretty large physical radius included in the same coarse-graining element, which shrinks as the energy/mass becomes higher and one approaches the geometric limit.  


 \paragraph*{Decoherence in this paradigm.}

 In most instances, according to \eqref{equ:new_deco_path}, decoherence will follow from a large difference between the amplitude contributions of the different sets, e.g. $W_\alpha:=\int_{\gamma_{i}\in\alpha_i}  \mathcal{D}\gamma_{i}(\tau)e^{i(S(\gamma_{i}(\tau)))/\hbar}$ being such that $|W_\alpha|/|W_{\alpha\rq{}}|<<1$. This is enough to explain how obtaining \lq\lq{}which-path\rq\rq{} (welcher-weg)  information in  semi-classical interference experiments destroys interference. \lq\lq{}Which-path\rq\rq{}  information makes it so that the  sets of semi-classical paths no longer reach the same final configurations. Lack of interference from a given $W_\alpha$ will also come about, according to \eqref{equ:semi_field_interference}, if the corresponding Van-Vleck determinant is small, meaning that $\gamma^\alpha_{\mbox{\tiny{cl}}}$ reaching this final point requires a lot of fine-tuning of its initial direction in configuration space, according to \eqref{mom_VV}.

  As a simple example, let us use a single free particle on the unit radius $S^1$. I will use this to mimic the setup of a simple beam-splitter experiment, where the given particle has two options to go from an initial source to final detectors.  In this case, configuration space is given by a single variable, $\theta\in S^1$. In principle this would make a better relational analogy with an extra \lq{}time\rq{} configuration variable, $t$. The full analytic solution for that path integral over a ring can be found in the book by Schullman \cite{Schulman}. But the simplified example suffices for my purposes. Also note that I don\rq{}t want  higher winding numbers to play any role, supposing the particle to be emitted and absorbed at initial and (resp.) final points, although it would be easy to relax this condition.
  
        The action is given by 
 \be\label{circle1}S(\gamma)=\frac{1}{2}\int d\theta\,  \dot \gamma^2(\theta)\ee
 where $\dot \bullet= \frac{d\bullet}{d\theta}$,  and the equations of motion are $\ddot \gamma(\theta)=0$, meaning $\gamma_{\mbox{\tiny cl}}(\theta)=a\theta+b$, for $a, b$  constants. Suppose we want to look at solutions which  start at $\theta_i$ and end at $\theta_f$. Then it is easy to see there are two possible unit-parametrized solutions to the equations of motion interpolating between these configurations, i.e. such that $\gamma(0)=\theta_i$ and $\gamma(1)=\theta_f$:  $\gamma^1_{\mbox{\tiny cl}}(t)=\theta_f t+(1-t)\theta_i$  and $\gamma^2_{\mbox{\tiny cl}}(\theta)= (\theta_f-2\pi)t+(1-t)\theta_i$. Thus the on-shell actions are:  
 \be\label{circle2}S(\gamma^1_{\mbox{\tiny cl}})=\frac{1}{2}\int_0^1 (\theta_f-\theta_i)^2d\theta~~ \mbox{and}~~S(\gamma^2_{\mbox{\tiny cl}})=\frac{1}{2}\int_0^1 ((\theta_f-2\pi)-\theta_i)^2d\theta\ee
 the Van Vleck determinants give
 \be\label{circle3}\det{\left|\frac{\delta^2 S(\gamma^1_{\mbox{\tiny cl}})(\theta_i,\theta_f)}{\delta\theta_i\delta\theta_f}\right|}=|-\frac12|=|\frac12|=\det{\left|\frac{\delta^2 S(\gamma^2_{\mbox{\tiny cl}})(\theta_i,\theta_f)}{\delta\theta_i\delta\theta_f}\right|}\ee
 which was to be expected, as there is no room for paths to expand into.  Thus, according to \eqref{equ:semi_field_interference}, there will be interference. For example, for the symmetric choice $\theta_i=0$ and $\theta_f=\pi$, for  $\theta\in[-\pi,\pi]$, where $\pi$ is identified with $-\pi$, as usual, still from \eqref{equ:semi_field_interference} we get complete constructive interference with $S(\gamma^1_{\mbox{\tiny cl}})=S(\gamma^2_{\mbox{\tiny cl}})$ in \eqref{circle2}.
   \begin{figure}[h]
\begin{center}
\includegraphics[width=0.75\textwidth]{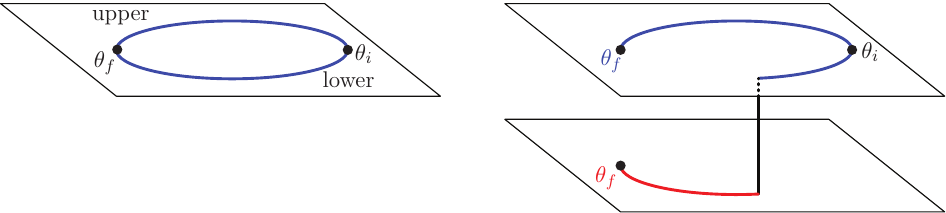}
\caption{The two setups: on the lhs,  the original one, an upper and a lower path robustly interfere. On the rhs, a second setup, which has \lq{}which-path\rq{}  information, translating one of the extremal trajectories in configuration space. The vertical displacement has been colored to represent the different parts of configuration space each path is exploring.  }
\end{center}\label{fig:ring}
\end{figure}   
 
  To describe the which-path information, suppose  that configuration space is two copies of $S^1$, i.e. $S^1\times {weg}$ where  ${weg}=\{0,1\}$ (for welcher-weg). We could have set $\{0\}$=\rq{}red\rq{} and $\{1\}$=\rq{}blue\rq{} as two alternative tags of the particle. Now, we can model something that indicates the passage of the path through a given point in $S^1$ by an action valued in $\{0,1\}$. We want the classical paths to register whether they have passed through a given point -- let\rq{}s say $\theta=\pi/4$ -- where the detector is placed in $S^1$, i.e. we would like e.g.: 
 \be\label{equ:which-path} \gamma^1_{\mbox{\tiny cl}}(t)=(\theta_f t+(1-t)\theta_i\,, \,\Theta[\pi/4-(\theta_f t+(1-t)\theta_i)] )
 \ee similarly for $\gamma^2_{\mbox{\tiny cl}}(t)$,   where we have written the curve component-wise in $S^1\times {weg}$ where, again,  ${weg}=\{0,1\}$, and $\Theta$ is the Heaviside step function. The Heaviside can be obtained as the derivative of the ramp function, $R(x)=\frac{x+|x|}{2}$, which can itself be easily integrated,  and it is straightforward to find the new action that corresponds to classical curves \eqref{equ:which-path} from the previous \eqref{circle1}-\eqref{circle3}, including their identical Van-Vleck determinant (for the same reasons). Notice that the \lq{}detector\rq\rq{}  jumps from 0 to 1 if the classical path ever goes through $\pi/4$, as required.

 Now if the initial point is $(\theta_i, weg)=(0,0)$ and the final is $(\theta_f, weg)=(\pi,0)$, unlike the previous case, there is only one path that extremizes the action, $\gamma^1_{\mbox{\tiny cl}}(t)=(\theta_f t+(1-t)\theta_i\,, \,\Theta[\pi/4-(\theta_f t+(1-t)\theta_i)] )$, and no significant contribution from paths with initial positive angular velocity to the final amplitude (see figure 1). Thus there is no significant interference effect.\footnote{ In this simple example, the action was not presumed to be given by the usual length functional on configuration space, as is \eqref{circle1}. However, if it was, and the distance $d((\pi,0), (\pi,1))>>\hbar$, there would still be negligible interference.} Replacing $weg$ with a \lq\lq{}bomb\rq\rq{}  as in  the Elitzur–Vaidman bomb-testing \cite{bomb}, it is easy to see that in the presence of such a \lq{}bomb\rq{}, there is never semi-classical interference between the two paths,  \textit{even though the projection to $S^1$  looks identical to the previous case}, when we look at the entire configuration space, we can see  the two alternatives end up in different points in configuration space.  The confusion arises if we restrict attention to the trajectories in physical space, and not to configuration space. This lack of interference effect from two alternative coarse-grainings is a simplified model for the present notion of decoherence in configuration space.

\section{Records}\label{sec:records}
As originally argued in \cite{Page_Wootters, Page_summary},  \begin{quote}
\lq\lq{}We cannot compare things at different times, but only different records at the same time. We cannot know the past except through its records in the present, so it is only present records that we can really test. [...] For example, we cannot directly test the probability that an electron has spin up at $t=t_f$ given that it had spin up at $t=t_i< t_f$, but only given that there are records at $t=t_f$ that we interpret as indicating the electron had spin up at $t=t_i$.\rq\rq{} (Page, \cite{Page_summary})
\end{quote}


Even if there is some objective meaning to a relational transition amplitude, $W(\phi_i, \phi_f)$, given the Past Hypothesis, we should be able to do science from assumptions about the frequency of configurations in $\mathcal{Q}$. In fact, what we do have access to right before we examine the results of an experiment are the memories, or records, of the setup of the experiment. This is what Carroll and Sebens call the `post-measurent, pre-observation' state of affairs \cite{Carroll}.  These memories are somehow encoded in the present configuration. For example, they can be presently encoded in in the form of  a rock formation, pencil markings on a book, the setup of instruments in a laboratory, a photograph,  a specific neural circuit in place, etc. What we do, is hold some of these records fixed, and compare with the presence of some other properties of our configuration. It is the consistency between records that gives us a way to infer the laws of Nature and to do Science.

 Since there is no absolute time in the picture, or space-time, the habitat of measurements and the updating of probabilities is significantly modified.  My addition to the literature in this respect is to give a semi-classical definition of the mathematical structure to be expected of such records, and why they effectively function as a \lq{}past measurement\rq{}. 



\subsection{Semi-classical records}\label{sec:sc_records}
  I will denote a record-holding subset of $\mathcal{Q}$ as follows: if the record is $\phi_r$,  the record-holding manifold is  $\mathcal{Q}_{(r)}$. Each copy of the experimenter coexisting with the given record  will find itself in a specific configuration $\phi\in \mathcal{Q}_{(r)}$. The pre-selection in an experimental setting is the selection of the manifold $\mathcal{Q}_{(r)}$, consisting of all those configurations with the same records. The post-selection is the finding out of where in that set your own configuration is.

 I will then properly define a configuration $\phi$ as holding a  \emph{semi-classical record} of a configuration $\phi_r$ if it obeys the following criterion: for a given  `in' configuration ${\phi^*}$, all of the elements of the minimal PEC's  have to contain the `recorded' configuration $\phi_r$ (i.e. the configuration which $\phi$ \emph{holds a record} of).  This is meant to give the intuitive notion that $\phi_r$'s ``happening" is encoded in the state $\phi$, i.e. at least semi-classically, `branches' of the wavefunction contributing to the amplitude of $\phi$ can't `skip' $\phi_r$. I thus define: 
\begin{defi}[Semi-classical record  (single segmented)]\label{def:full_record}
Given an initial configuration ${\phi^*}$, $\phi$ and  $\{C_\alpha\}_{\alpha\in I}$ the extremal coarse graining (def. \ref{def:ECS}) between $\phi^*$ and $\phi$,  of radii $\rho^\alpha_{\mbox{\tiny max}}$ given in \eqref{rhomax}, we will say  $\phi$ holds  a (single-segmented) semi-classical record of a field configuration $\phi_r$, if the ball $B_{\rho_{\mbox{\tiny{max}}}}(\phi_r)$ is contained in every $C_\alpha$. (see figure 2). 
\end{defi}
 Although I have here presented the definition using ECs, for geodesics which are not segmented (single segmented), an identically formulated, but  more general, $n$-segmented definition using  PEC  is straightforward (see definition \ref{def:record_PEC} in the appendix).
 \begin{figure}[h]
\begin{center}
\includegraphics[width=0.75\textwidth]{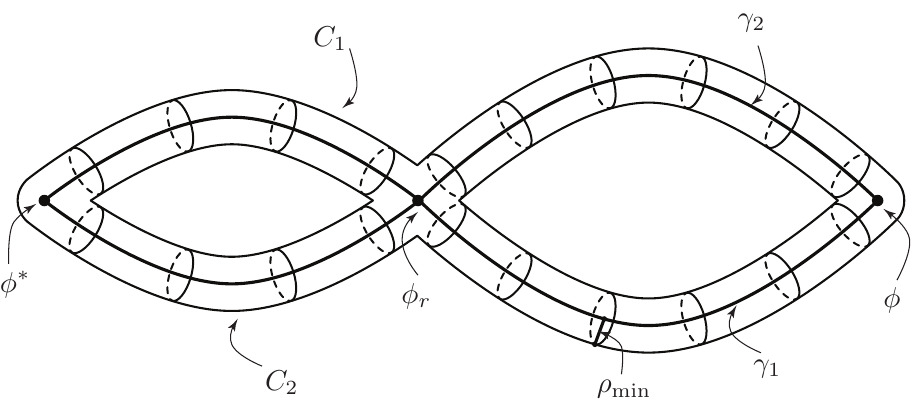}
\caption{An extremal coarse-graining between $\phi^*$ and $\phi$ in configuration space $\mathcal{Q}$, consisting of the elements $C_1$ and $C_2$, seeded by the extremal paths $\gamma_1, \gamma_2$, with radius $\rho_{\mbox{\tiny max}}$. Here $\phi$ contains a single segmented semi-classical record of $\phi_r$.}
\end{center}\label{fig:record}
\end{figure}   
It is also worth to note that,  up to this order of approximation, our coarse-graining cannot resolve points through the transition amplitude.  Therefore a better visualization of a record would actually be the ball, $B_{\rho_{\mbox{\tiny{max}}}}(\phi_r)$. For this reason, I chose this 'hybrid definition' in terms of $C_\alpha$,  $B_{\rho_{\mbox{\tiny{max}}}}(\phi_r)$, and the configurations $\phi,  \phi^*$.  I could have similarly defined records only in terms of balls around $\phi^*, \phi, \phi_r$, and $C_\alpha$, or in terms of $C_\alpha$ and only the configurations $\phi, \phi_r, \phi$ (since $C_\alpha$ already limits our access to these points to the balls around them).   I judged this hybrid formulation most informative. Also note that, to prove our assertions below, I will need the precise properties of the semi-classical approximations, which require the particular properties of the representative configuration of the record, $\phi_r$.  For simplicity I have called this representative \lq\lq{}the semi-classical record".

With this definition I will prove that: 
\begin{theo}\label{theo:record}
Given a configuration $\phi$ with a semi-classical record of $\phi_r$ to order $\hbar$, then
\be\label{equ:semi_classical_record} W_{{\mbox{\tiny{cl}}}}({\phi^*},\phi)=  W_{{\mbox{\tiny{cl}}}}({\phi^*},  \phi_r)W_{{\mbox{\tiny{cl}}}}( \phi_r,  \phi)+\order{\hbar^2}
  \ee
Given the factorization property \eqref{equ:factorization} for the density functional, i.e. $F(z_1z_2)=F(z_1)F(z_2)$, it means that the equation for the density of $\phi$ automatically becomes an equation for conditional probability on $\phi_r$, 
\be
\label{equ:conditional} \mu(\phi)=\mu(\phi_r)\mu(\phi|\phi_r)
\ee  where $\mu(\phi|\phi_r)=|W(\phi_r,\phi)|^2$.  
\end{theo}
We first note that for the semi-classical record to be of order $\hbar$,   the length of the extremal paths that seed the coarse-graining, $\gamma_\alpha^{\mbox{\tiny{cl}}}$, need to obey  $S_\alpha:=S(\gamma_\alpha^{\mbox{\tiny{cl}}})>>\hbar$.   Thus this coarse-graining allows us to use the semi-classical approximation  \eqref{equ:semi_classical_exp}, 
$$ W_{{\mbox{\tiny{cl}}}}({\phi^*}, \phi)=  \sum_{\gamma_{\mbox{\tiny{cl}}}}(\Delta_{\gamma_{\mbox{\tiny{cl}}}})^{1/2}\exp{\left(i S_{\gamma_{\mbox{\tiny{cl}}}}({\phi^*},\phi)/\hbar\right)}+\order{\hbar^2}$$
If the system was deparametrizable,\footnote{And quadratic in momenta so that we could write the path integral solely in configuration space.} we could use \eqref{equ:deparametrizable},  and the semi-classical composition law \cite{Kleinert} to write: 
\be\label{sc_non_r} W_{{\mbox{\tiny{cl}}}}({\phi^*},\phi)= \int \mathcal{D}\bar\phi_m W_{{\mbox{\tiny{cl}}}}((t_*,{\bar\phi^*}),  (t_m,\bar\phi_m))W_{{\mbox{\tiny{cl}}}}( (t_m,\bar\phi_m),  (t,\bar\phi))
\ee
for an intermediary time $t_m$.  Choosing $t_m=t_r$, the integral gains a  $\delta(\bar\phi_r)$, since the extremal paths  at $t_r$ pass uniquely through $\bar\phi_r$, and we obtain the theorem:
$$ W_{{\mbox{\tiny{cl}}}}({\phi^*},\phi)=  W_{{\mbox{\tiny{cl}}}}((t_*,{\bar\phi^*}),  (t_r,\bar\phi_r))W_{{\mbox{\tiny{cl}}}}( (t_r,\bar\phi_r),  (t,\bar\phi))
$$
 In the more general case however, we must do some work but the gist of the proof is similar.

One can still use the semi-classical limit for obtaining an effective parametrization along each path. One can find a composition law from the on-shell action functionals themselves, without reference to time. The strategy for the proof is to first obtain the semi-classical integration composition properties by the standard steepest descent Gaussian measure.  But because reparametrization of the curves act trivially,  one must find a replacement for what a 'constant time surface' represents; we will use this surface to integrate perturbations. In a diagram that contains space and time, an infinitesimal constant time surface is after all just a codim=1 transversal surface to the trajectory. Here, we need such a surface to also coincide for all the extremal paths. 
 
Suppose there exists a single extremal path between $\phi^*$ and $\phi$. Consider the integral: 
  \be
 \int \mathcal{D}\phi_m \Delta_1^{1/2}\exp{\left(i S({\phi^*},\phi_m)/\hbar\right)}\Delta_2^{1/2}\exp{\left(iS(\phi_m,\phi_f)/\hbar\right)} 
  \ee
  where the integral should be evaluated by stationary phase. One issue we face straightaway is that because of reparametrization invariance of the extremal path, there is a redundancy for the stationary value $\phi_m=\tilde\phi_m$ being placed along the extremal path. Given an arbitrary auxiliary parametrization of the extremal path, $\gamma_{{\mbox{\tiny{cl}}}}(\tau)$ we can gauge-fix the integral $ \int \mathcal{D}\phi_m$ to take place in a transversal plane to $\frac{d}{d\tau}\gamma_{{\mbox{\tiny{cl}}}}(\tau)$, obtaining an overall factor $\int \mathcal{D}\tau=M$: 
   \begin{eqnarray}
 &~&\int \mathcal{D}\phi_m \Delta_1^{1/2}\exp{\left(i S({\phi^*},\phi_m)/\hbar\right)}\Delta_2^{1/2}\exp{\left(iS(\phi_m,\phi_f)/\hbar\right)}\nonumber\\
  &=&M  \int \mathcal{D}\phi^{\perp}_{tm} \Delta_1^{1/2}\exp{\left(i S({\phi^*},{\phi^\perp_{tm}})/\hbar\right)}\Delta_2^{1/2}\exp{\left(iS({\phi^\perp_{tm}},\phi_f)/\hbar\right)}
  \end{eqnarray}
Where the variables $\phi^\perp_{tm}$ now span the transversal plane to $\phi_{\mbox{\tiny cl}}$ at some $\tau=t$ which we will discuss below.   I have to show that no field-dependent Fadeev-Popov determinant arises from such gauge-fixing. For that, suppose that there exists a a gauge-invariant metric in configuration space, \lq{}$\cdot$\rq{}.\footnote{This is not true for general relativity in ADM  form \cite{ADM}, which possesses refoliation symmetry. But it does hold for conformal geometrodynamic theories, such as shape dynamics, with an inner product of the form $$\langle u, v\rangle_g=\int_M
\sqrt{g}d^3x \, \, \sqrt{C^{ab} C_{ab}}\,\,u_{ab}g^{ac}g^{bd} v_{cd}$$ for $C_{ab}$ the Cotton tensor, and $u,v\in T_g\mathcal{Q}$. And the argument here goes through without a hitch \cite{Conformal_geodesic}.}
   Then, we  gauge-fix the reparametrizations, by arc-length: 
 \be\label{equ:arc_length}
G_\tau:=\sqrt{\gamma\rq{}\cdot \gamma\rq{}}-1=0
 \ee
 The infinitesimal version, i.e. for $\delta_\epsilon \tau= c+\epsilon \tau+\order{\epsilon^2}$, 
 $$G_{\tau^\epsilon}= |\epsilon|\sqrt{\gamma\rq{}\cdot \gamma\rq{}}-1$$
gives the variation: 
\be \label{equ:FP_time}
\frac{\delta G_{\tau^\epsilon}}{\delta\epsilon}_{|G_\tau=0} = \pm\sqrt{\gamma\rq{}\cdot \gamma\rq{}}_{|G_\tau=0}=\pm 1
\ee
And thus we fix $\tau$ as being given by arc-length $t$ along the extremal path. 
    
   The integral is being evaluated by stationary phase, so in addition to evaluating the functions multiplying the exponential at the stationary phase point, i.e. 
   ${{\phi^\perp_{tm}}} = \tilde\phi_{tm}$, we also have to find the steepest descent contribution of the Gaussian integration around the critical point. If we call the exponent of the product of exponentials $T$, then we want to Taylor expand this exponent function around the critical point to second order in the position variable ${{\phi^\perp_{tm}}}$. The zeroth order is a phase that we build into our phase convention, the first order vanishes at the critical point, and the second order is given by
\be\label{T2}T^{(2)} =\left(\frac{\delta^2 S({\phi^*},{{\phi^\perp_{tm}}})}{\delta{\phi^{\perp2}_{tm}}}+\frac{\delta^2 S({{\phi^\perp_{tm}}}, \phi)}{\delta{\phi^{\perp2}_{tm}}}\right) ({\phi^{\perp}_{tm}}-\tilde\phi_{tm})^2\ee
Performing this Gaussian integral then gives us the inverse of the square root of this pre-factor, $$\det{\left(\frac{\delta^2S(\phi^*,{\phi^\perp_{tm}})}{\delta{\phi^{\perp2}_{tm}}}+\frac{\delta^2S({\phi^\perp_{tm}},\phi)}{\delta{\phi^{\perp2}_{tm}}}\right)}^{-1}.$$ 
which will be absorbed by the semi-classical integral measure, $\mathcal{D}\phi^\perp_{tm}$.

  Thus we need to show that the Van-Vleck determinant satisfies the correct gluing property:
\be\label{equ:VV_comp}\det{\left(-\frac{\delta^2S(\phi^*,{\phi^\perp_{tm}})}{\delta\phi^*\delta{\phi^\perp_{tm}}}\right)}\det{\left(-{\frac{\delta^2S({\phi^\perp_{tm}},\phi)}{\delta{\phi^\perp_{tm}}\delta\phi}}\right)}\det{\left(\frac{\delta^2S(\phi^*,{\phi^\perp_{tm}})}{\delta{\phi^{\perp 2}_{tm}}}+\frac{\delta^2S({\phi^\perp_{tm}},\phi)}{\delta{\phi^{\perp2}_{tm}}}\right)}^{-1}
=\det{\left(-\frac{\delta^2S(\phi^*,\phi)}{\delta\phi^*\delta\phi}\right)}\ee
which is the required composition law of determinants.

The action is additive, thus for the given extremal path going through the intermediary ${\phi^\perp_{tm}}$
\be\label{equ:additive}
S_{\gamma_{\mbox{\tiny{cl}}}}({\phi^*},\phi)=S_{\gamma^1_{\mbox{\tiny{cl}}}}({\phi^*},{\phi^\perp_{tm}})+S_{\gamma^2_{\mbox{\tiny{cl}}}}({\phi^\perp_{tm}},\phi)
\ee
Using \eqref{equ:additive}, from the  stationarity condition at an intermediary field configuration, we have (dropping the subscripts for clarity):
\be\label{equ:intermediary_condition}\frac{\delta S(\phi^*,{\phi^\perp_{tm}})}{\delta {\phi^\perp_{tm}}}+\frac{\delta S({\phi^\perp_{tm}},\phi_f)}{\delta {\phi^\perp_{tm}}}=-\pi_f(\phi^*,{\phi^\perp_{tm}})+\pi_i({\phi^\perp_{tm}},\phi)=0
\ee this requires the momenta to be continuous at ${{\phi^\perp_{tm}}}$, setting ${{\phi^\perp_{tm}}}$ to be along the extremal path ${{\phi^\perp_{tm}}}=\tilde\phi_{tm}$. Thus, given a classical path $\gamma$,  ${{\phi^\perp_{tm}}}$ also depends on $\phi$ and $\phi^*$. It changes in accordance to changes in $\phi$ so that equation \eqref{equ:intermediary_condition} remains true. Thus, deriving \eqref{equ:intermediary_condition} by $\phi$:
$$-\frac{\delta\pi_f(\phi^*,{{\phi^\perp_{tm}}})}{\delta{\phi^\perp_{tm}}}\frac{\delta{\phi^\perp_{tm}}}{\delta\phi}+\frac{\delta\pi_i({{\phi^\perp_{tm}}},\phi)}{\delta{\phi^\perp_{tm}}}\frac{\delta{\phi^\perp_{tm}}}{\delta\phi}+\frac{\delta\pi_i({{\phi^\perp_{tm}}},\phi)}{\delta\phi}=0
$$ 
\be\label{1}\Rightarrow~~-\frac{\delta\pi_f(\phi^*,{{\phi^\perp_{tm}}})}{\delta\phi^\perp_{tm}}+\frac{\delta\pi_i({{\phi^\perp_{tm}}},\phi)}{\delta{\phi^\perp_{tm}}}=-\frac{\delta\pi_i({{\phi^\perp_{tm}}},\phi)}{\delta\phi}\left(\frac{\delta{\phi^\perp_{tm}}}{\delta\phi}\right)^{-1}
\ee
Writing equation  \eqref{equ:VV_comp} in terms of momenta we have: 
\be\label{equ:intermediary} 
\det{\left(-\frac{\delta \pi_i(\phi^*,{{\phi^\perp_{tm}}})}{\delta\phi^\perp_{tm}}\right)}\det{\left(-{\frac{\delta\pi_f({{\phi^\perp_{tm}}},\phi)}{\delta\phi^\perp_{tm}}}\right)}\det{\left(-\frac{\delta\pi_f(\phi^*,{{\phi^\perp_{tm}}})}{\delta\phi^\perp_{tm}}+\frac{\delta\pi_i({{\phi^\perp_{tm}}},\phi)}{\delta\phi^\perp_{tm}}\right)}^{-1}=\det{\left(-\frac{\delta\pi_i(\phi^*,\phi)}{\delta\phi}\right)}
\ee

Finally, substituting   \eqref{1} into the lhs of arguments of the deteminants in \eqref{equ:intermediary}:
\begin{align} 
& {\left(-\frac{\delta \pi_i(\phi^*,{{\phi^\perp_{tm}}})}{\delta{\phi^\perp_{tm}}}\right)}{\left(-{\frac{\delta\pi_f({{\phi^\perp_{tm}}},\phi)}{\delta{\phi^\perp_{tm}}}}\right)}
\left(-\frac{\delta\pi_i({{\phi^\perp_{tm}}},\phi)}{\delta\phi}\left(\frac{\delta\phi^\perp_{tm}}{\delta\phi}\right)^{-1}\right)^{-1}\nonumber\\
=&
\left(-\frac{\delta \pi_i(\phi^*,{\phi^\perp_{tm}})}{\delta{\phi^\perp_{tm}}}\right)\left(\frac{\delta{\phi^\perp_{tm}}}{\delta\phi}\right)\left(-{\frac{\delta\pi_f({{\phi^\perp_{tm}}},\phi)}{\delta{\phi^\perp_{tm}}}}\right)\left(-\frac{\delta\pi_i({{\phi^\perp_{tm}}},\phi)}{\delta\phi}\right)^{-1}\nonumber\\
=&{\left(- \frac{\delta \pi_i(\phi^*,{\phi})}{\delta\phi}\right)} \end{align}
where we note that an extra cancellation occurs since 
$\frac{\delta \pi_i({{\phi^\perp_{tm}}},\phi)}{\delta\phi}=\frac{\delta \pi_f({{\phi^\perp_{tm}}},\phi)}{\delta\phi^\perp_{tm}}$ because  they are equal  as second derivatives of the action. Thus  \eqref{equ:VV_comp} is proven. 

Thus for this single extremal path, $\gamma=\gamma_{1}\circ\gamma_{2}$ we obtain:
\be
\int \mathcal{D}\phi^{\perp}_{tm} \Delta_1^{1/2}\Delta_2^{1/2}\exp{\frac{i}{\hbar}\left( S({\phi^*},{\phi^\perp_{tm}})+S({\phi^\perp_{tm}},\phi)\right)}\approx \Delta^{1/2}_{\gamma}\exp{iS(\phi^*, \phi)/\hbar}
\ee
Moreover, since there exists only one extremal point in the transversal plane, the integral is peaked around $\phi^\perp_{tm}-\tilde\phi_{tm}$, i.e. we have approximately\footnote{The riddance of $T^{(2)}$ of \eqref{T2} in this expression can be seen as a Jacobian coming from  fixing the variations along the transversal planes, \textit{at the base space point} of the double-tangent bundle: $\phi^\perp_{tm}=\tilde\phi_{tm}$. } 
\be
 \Delta_1^{1/2}\exp{\left(i S({\phi^*},\tilde\phi_{tm})/\hbar\right)}\Delta_2^{1/2}\exp{\left(iS(\tilde\phi_{tm},\phi)/\hbar\right)}\approx   \Delta^{1/2}\exp{\left(i S({\phi^*},\tilde\phi)/\hbar\right)}
\ee where the variations in $\Delta_1, \Delta_2$ are transversal to the extremal path. 

  Each overall extremal path decomposes into a unique composition  of extremal paths,  $\gamma_{\alpha}=\gamma_{\alpha_1}\circ\gamma_{\alpha_2}$ with final (resp. initial) endpoints on $\tilde\phi_{tm}=\phi_r$. Now, according to our assumptions, the only point that all of the extremal paths share is $\phi_r$. More than that, all the tubular bundles include the ball $B_{\rho}(\phi_r)$.
  
    It is also important to know that in all infinite-dimensional Riemannian spaces, and generically even in non-compact finite-dimensional spaces, the geodesics at the cut-loci always admit a codim=1 transversal plane \cite{Piccione}.\footnote{This is not true for orbifolds (which does not mean the formula doesn\rq{}t hold, it just means this proof needs to be amended). Also note that it is not necessary for this proof that all of the curves define the same orientation in the codim=1 surface. I.e. it is not necessary that the define the same \lq\lq{}direction of time\rq\rq{}. } I will assume that this is true more generally of generic dynamical systems in infinite-dimensional configuration spaces.
  
    Therefore choosing a common transversal plane $\phi_{r}^\perp$ for all extremal paths --- which is one of the roles an equal-time surface plays--- we note that this transversal plane furthermore fills a section of each of the tubular bundles $C_\alpha$, since it intersects the ball $B_{\rho_{\mbox{\tiny{max}}}}(\phi_r)$ which is contained in each $C_\alpha$. Putting it all together, we obtain:
\be\label{equ:semi_classical_comp}  \sum_{\alpha_1}\Delta_{\alpha_1}^{1/2}\exp{\left(i S_{\alpha_1}({\phi^*},\phi_r)/\hbar\right)}\sum_{\alpha_2}\Delta_{\alpha_2}^{1/2}\exp{\left(iS_{\alpha_2}(\phi_r,\phi_f)/\hbar\right)} = \sum_{\alpha}\Delta_{\alpha}^{1/2}\exp{\left(iS_{\alpha}(\phi^*,\phi)/\hbar\right)}
\ee
up to orders $ \mathcal{O}({\hbar^2}$).  $\square$\vspace{.3cm}

It is these two composition properties which ensure that the leading order of the multiple sum (implicit in each kernel) in the lhs of \eqref{equ:semi_classical_comp} reduce to a single sum on the rhs. In the particle case the last equality of equation \eqref{equ:semi_classical_comp} is a consequence of the fact that  the intermediary integration is at fixed time -- a time which in our case is included in the configuration, and so we had to find an alternative procedure. Nonetheless, the results are that only one representative of each path can be taken. 
\begin{figure}[h]
\begin{center}
\includegraphics[width=0.4\textwidth]{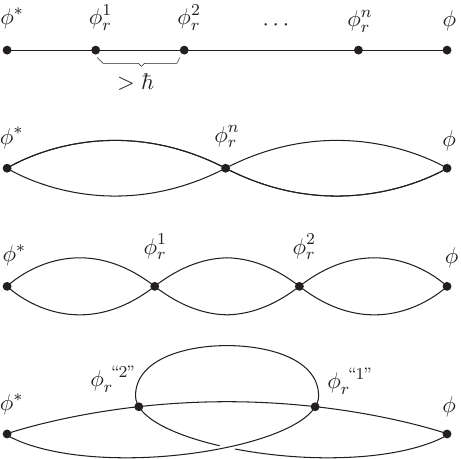}
\caption{The types of sequence of records that can arise. The first gives rise to a notion of a granular history. The second and third represent a more general version of records of records. The fourth represents the sort of records which are not consistent, as their ordering disagrees. }
\end{center}\label{fig:screens}
\end{figure} 
 We can compare transition amplitudes for configurations with the same records in the following way, given $\phi_1, \phi_2 \in \mathcal{Q}_{(r)}$, for any initial ${\phi^*}$:
\be\label{equ:relative_prob}
\frac{\mu({\phi^*}, \phi_1)}{\mu({\phi^*}, \phi_2)}\approx \frac{\mu({\phi^*},\phi_r)\mu(\phi_r, \phi_1)}{\mu({\phi^*},\phi_r)\mu(\phi_r, \phi_2)}= \frac{\mu(\phi_r, \phi_1)}{\mu(\phi_r, \phi_2)}
\ee
This ratio gets rid of a common quantity to both density functions; carrying away the relevance of how the system actually got to that record in the first place (except as in restricting the possible initial conditions of the system itself). 

\paragraph*{Strings of records} 
 
 The definition of our preferred coarse-grainings disallows self-interesection of any single given element of the coarse-graining. This implies that, for e.g.  two records  $\phi_r^1, \phi_r^2$, the  extremal path $\gamma_\alpha\in \Gamma({\phi^*},\phi)$ must pass through  $\phi_r^1, \phi_r^2$ in a given order. That is, if   $\phi\in \bigcap_i\mathcal{Q}_{(r^i)}$ ($\phi$ contains multiple semi-classical records $\phi_r^i$), an ordering of the records exists for \emph{each} element of the coarse-graining, meaning we can rewrite, for each $\alpha$, the set $\{\phi_r^i\}_{i=1\cdots n}$ as an ordered set $(\phi_r^1,\, \cdots\,,\,\phi_r^n)$. 
 
  If there is just one coarse-graining, i.e. just one element $\alpha$, and the on-shell action of the extremal path seed $\gamma_\alpha$ between $\phi_r^i$ and $\phi^{i+1}_r$ is much larger than $\hbar$, this implies that $P_{{\mbox{\tiny{cl}}}}({\phi^*},\phi^{i+1}_r)=P_{{\mbox{\tiny{cl}}}}({\phi^*},\phi^{i}_r) P_{{\mbox{\tiny{cl}}}}({\phi^i_r},\phi^{i+1}_r)$. Thus the density  decomposes:
 $$P_{{\mbox{\tiny{cl}}}}({\phi^*},\phi)\approx  P_{{\mbox{\tiny{cl}}}}({\phi^*},  \phi_{r}^{i})P_{{\mbox{\tiny{cl}}}}(\phi_{r}^i,  \phi_{r}^{i+1})P_{{\mbox{\tiny{cl}}}}( \phi_{r}^{i+1},  \phi)
 $$ 
 In the simple case with a single $\alpha$, the records of $\phi\in \mathcal{Q}_{(r)}$ coincide with a coarse-grained classical history of the field,  in this case \emph{we recover a notion compatible with classical Time}.  This is the first case represented in figure 3. 
  In the more general case of there being more than one element of the coarse-graining, we will have  a consistent string of records if the ordering $(\phi_r^{\alpha_1},\, \cdots\,,\,\phi_r^{\alpha_n})$ given for each $\alpha$, coincide among all the different $\alpha$, i.e. $\phi_r^{\alpha_i}=\phi_r^{\alpha'_i}=\phi_r^ i$ for all $i$ and any $\alpha, \alpha'$ elements of the coarse-graining. These are the two following cases in figure 3. We can once again decompose the kernel:    
  \be\label{equ:records_string}
  P_{{\mbox{\tiny{cl}}}}({\phi^*},\phi)\approx  P_{{\mbox{\tiny{cl}}}}({\phi^*},  \phi^1_r)P_{{\mbox{\tiny{cl}}}}( \phi^1_r,  \phi_r^2)\cdots P_{{\mbox{\tiny{cl}}}}( \phi^n_r,  \phi)
  \ee  Thus, even if there are no records regarding which slit an electron went through in a double-slit experiment, we can still have records of the setup of the experiment and its results on the screen.

    If, on the other hand, the ordering does not coincide among the different $\alpha$, there is no absolute ordering relation $\prec$ between the $\phi_r^i$, and equation \eqref{equ:records_string} does not hold. {Equation \eqref{equ:records_string} will hold if the system is deparametrizable, as in \eqref{equ:deparametrizable}, but we can have weaker conditions.  In fact,  equation \eqref{equ:records_string} will hold if there exist  consistent families of \emph{`clock-like' variables} in the given region of configuration space possessing the records. I am here calling a clock-like variable on a given region $\mathcal{U}$ a scalar functional $T[\phi]$ of the configuration variables that is smooth, for which $\delta T_{|\phi_r}\neq 0$, and which is strictly monotonic along extremal paths in that region. More explicitly, for $\phi_1, \phi_2\in \mathcal{U}$ a given region $\mathcal{U}\subset\mathcal{Q}$, then $T\in C(\mathcal{U})$ is a clock-like functional if, for $\phi(t)\in\Gamma(\phi_1, \phi_2)\cap \mathcal{U}$ an extremal curve in region $\mathcal{U}$, then $T[\phi(t)]$ is monotonic in the arbitrary time parametrization $t$ (i.e. either $T[\phi(t)]\geq  T[\phi(t')]$ for all $t\geq t'$ or $T[\phi(t)]\leq  T[\phi(t')]$ for all $t\geq t'$).\footnote{Note that arc-length, for example, does increase along each path, but it is not a scalar function in configuration space.} {For  example, the total volume of the Universe would be a choice of such a variable in regions of configuration space where classical dynamics shows  no recollapse.}  If there are two intersecting such regions $\mathcal{U},\,  \mathcal{U}'$, two clock-like variables  $T,\,  T'$ will be consistent if they have the same direction of monotonicity at the intersection. }  
 
\paragraph*{Another approach: coarse-graining of configuration space. An example.}

In the previous section, we had some labor to define coarse-grained histories and records. In the end, these coarse-grainings embody our limited access to configuration space. An alternative ---available at least when dealing with the configuration space of fields--- is to coarse-grain configuration space itself, not just the paths. One could use a cut-off on the field degrees of freedom, for example, on the eigenvalues of a Laplacian. To be more specific, suppose configuration space is the space of smooth sections of some tensor bundle $E$ over a closed spatial manifold $M$, e.g. $E=TM\otimes \cdots TM\otimes TM^*\otimes\cdots TM^*$, with  $\mathcal{Q}=C^\infty(E)$ being the space of smooth sections of this bundle.  This could be the space of positive (0,2) tensors, for which $\mathcal{Q}=C^\infty(E)$ would give the configuration space of gravity. In that case, given some homogeneous background, such as the round-sphere, (which here we have identified with $\eta=q^*$), we could use its associated Laplacian eigenfunctions $\nabla_\eta^2v_{ab}^i=\lambda_i v_{ab}^i$, $i\in I$, where $I$ is a countably infinite set, to form a \lq\lq{}coarse-grained configuration space\rq\rq{}, 
$$\mathcal{Q}_\lambda =\{ g_{ab}=\sum_i\alpha_i v_{ab}^i, \alpha_i \in \mathbb{R}, i\in I_{\mbox{\tiny fin}}\subset I\} 
$$
where $ I_{\mbox{\tiny fin}}$ is a finite subset of $I$ corresponding to an a UV regulator (for a closed manifold, there is no need of IR regulators). Such types of coarse-graining are only meaningful if they are accompanied by some dynamical decoupling of scales, as put forward in \cite{Locality_riem}, and usually assumed from the effective field theoretic framework.  

It is useful to keep this picture in mind, as it allows, in the following, allusion to trajectories in a coarse-grained configuration space, which is simpler than referring to properties of $C_\alpha$ at each turn. 
  
\subsection{Records and probability}\label{sec:Records and Bayes}
 There are many consequences to theorem \ref{theo:record}. First of all, it shows one of the things that we were aiming at: an emergent conditional probability equation, arising just from  the intrinsic configuration. It gives an ordering of configurations, one \lq\lq{}conditional\rq\rq{} on another, thus fulfilling one role of time; that of a (granular) linear ordering of instantaneous configurations.  This is clearest to see in the example with a single extremal path, the first of figure 3. There is another role of time, that of yielding \lq{}duration\rq{}, which I will not touch on here. Suffice it to say that such a notion needs to be extracted from the relative dynamics of subsystems. This is explored in \cite{QG_deco, Conformal_geodesic}. 
A similar decomposition should be available when finite-dimensional degrees of freedom decouple as well.

  \begin{figure}[h]
\begin{center}
\includegraphics[width=0.4\textwidth]{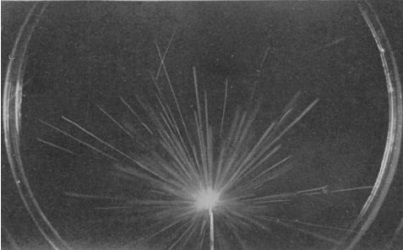}
\caption{A photograph of a bubble chamber experiment with atom decay. }
\end{center}\label{fig:bubble}
\end{figure}  
 An example of a system for which a linear ordering emerges in the same way is the Mott bubble chamber \cite{Mott}. In it, emitted particles from  $\alpha$-decay in a cloud chamber condense water vapor  along their trajectories (see figure 4). A quantum mechanical treatment involving a timeless Schroedinger equation finds that the wave-function peaks on configurations for which bubbles are formed collinearly with the source of the $\alpha$-decay. In this analogy, a `record holding configuration' would be any configuration with $n$ collinear condensed bubbles, and any configuration with $n'\leq n$ condensed bubbles along the same direction would be the respective  `record configuration'. In other words, the $n+1$-collinear bubbles configuration holds a record of the $n$-bubbles one.  For example, to leading order, the probability amplitude for  $n$ bubbles along the $\theta$ direction obeys \eqref{equ:records_string}:
\be\label{equ:Mott}P[(n,\theta),\cdots, (1,\theta)]\simeq P[(n\rq{},\theta),\cdots, (1,\theta)] P[(n\rq{},\theta),\cdots, (1,\theta)|(n,\theta),\cdots, (1,\theta)]\ee
 where $n\rq{}<n$, and $ P[B|A]$ is the conditional probability for $B$ given $A$.  
 
 One can in fact derive the probability \eqref{equ:Mott} from a simple detector model \cite{Halliwell_Mott}. Namely, for a a two-state string of detectors, with states given by $|0\rangle, |1\rangle$ and such that an interaction Hamiltonian between the $\alpha$ particles and the bubbles is given by: 
 $$ H_{\mbox{\tiny int}}=\sum f_k(q) (a_k+a_k^\dagger)$$ 
 where $f_k(q)$ is  localized in the region $B_k$ (representing the \lq{}bubble\rq{}),  and the states transform according to $a|0\rangle=0, a|1\rangle=|0\rangle, a^\dagger|0\rangle=|1\rangle, a^\dagger|1\rangle=0$. {In the case of the single particle, it is only one bubble, and one region, and configuration space matches physical space. However, the formalism could work similarly for many particles, and many bubbles, forming trajectories on a large dimensional configuration space.}
 
     Solving perturbatively (in the interaction parameter $\lambda$) for total Hamiltonian $H=H_o+\lambda H_{\mbox{\tiny int}}-E$, Halliwell obtains 
         the amplitude for $n$-bubbles to be excited, with the $n$-th bubble configuration being $q_f$, is given by:
 \be\label{Halliwell1}\langle q_f|\psi_n\rangle\propto \int d^Nq_n\cdots d^Nq_1 W(q_f,q_n)f_n(q_n)\cdots W(q_2,q_1)f_n(q_1)\ee
 where $W$ are propagators (for the free Hamiltonian), $f_i(q_i)$ are projections onto small regions of configuration space surrounding the $i$-th configuration, and $N$ is the dimension of configuration space. Clearly, if $f_k(q)\rightarrow \delta(q-q_k)$, one would regain a string of records, as in \eqref{equ:semi_classical_record}.\footnote{Of course, one should take into account that the definition \ref{def:full_record} of the record region itself, has some thickness, which does not require $f_k(q)\rightarrow \delta(q-q_k)$. See figure 2.}  In the standard semi-classical approximation \eqref{equ:semi_classical_exp} one obtains precisely, from \eqref{Halliwell1}: 
  $$\langle q_f|\psi_n\rangle\propto \int d^Nq_n\cdots d^Nq_1 \prod_{j=1}^n  \Delta^{1/2}(q_{j+1},q_j)f_j(q_j)\exp{\left(iS(q_{j+1},q_j)\right)}$$
 where $S(q_{j+1},q_j)$ is the on-shell action from $q_{j+1}$ to $q_j$. We then explicitly obtain our \eqref{equ:records_string} from $\langle q_f|\psi_n\rangle\overline{\langle q_f|\psi_n\rangle}$ (see the proof of \eqref{equ:semi_classical_record}). 
 
 For this derivation, Halliwell notes that it is essential that there is some asymmetry in configuration space, marked by the source of the $\alpha$-particles. It is the same here: we require  an\rq{}origin\rq{} of configuration space.

The definition implies that a  configuration can have many records, and an ordering among these, with earlier records being themselves recorded in later ones. This ordering of instantaneous configurations  from a fundamentally timeless theory gives rise to one  facet of Time.  Configurations that are far from  the given initial configuration $\phi^*$ -- but still connected to it by extremal paths -- will in general have more records, and concentrated amplitude. This ordering goes in line with the work of Barbour, Mercati and Koslowski \cite{Barbour_Arrow}, which suggests that complexity can give a good notion of an arrow of time for extremal paths in shape configuration space. Indeed it goes further, showing that complexity can be naturally related to ``time capsules" -- concentrations of wave-function amplitude along configurations with records \cite{Barbour94_2}.  

\paragraph*{Conservation of probability}
What we are talking about so far is volume in configuration space. How does that relate to probabilities, of the sort that is conserved? First of all, conserved in which \lq{}time\rq{}?  In the presence of a standard time parameter, we first distinguish between the total probability $P_t$ at one time, $t$, from $P_{t\rq{}}$ at another, $t\rq{}$.  Probabilities are usually taken to be conserved from one instant to another, and we must implement some similar notion of \lq{}surfaces\rq{} that relies solely on structures of configuration space; i.e. to translate this statement to one that uses only records and configuration space.

There can be (and generically there will be) an enormous amount of redundancy of records in $\mathcal{Q}_{(r)}$. In  other words, within $\mathcal{Q}_{(r)}$ there can be further, or redundant, record relations; $\phi_1, \phi_2\in \mathcal{Q}_{(r)}$  with $\phi_1\in \mathcal{Q}_{(\phi_2)}$. It is this redundancy  which we would not like to consider when discussing conservation of probabilities, and which therefore needs to be parsed.

A trivial example of such redundancy arises for strings of records \eqref{equ:records_string}. Consider subset of $\mathcal{Q}_{(r)}$ for which there is no such redundancy:
\be\label{equ:screen} \mathcal{S}_{(r)}:=\{\phi_i\in \mathcal{Q}_{(r)}~, i\in I\, |~ \phi_i\not\in \mathcal{Q}_{(j)}\, \, \forall i,j \in I\}
\ee
where I have abbreviated the record-holding submanifold of configuration $\phi_j$ as $\mathcal{Q}_{(j)}$ (see figure 5). I will call such a set \emph{a screen},  written as $\mathcal{S}_{(r)}\subset\mathcal{Q}_{(r)}$. 
 \begin{figure}[h]
\begin{center}
\includegraphics[width=0.4\textwidth]{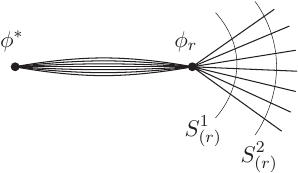}
\caption{A schematic representation of a sequence of two screens,  $S_{(r)}^1, S_{(r)}^2$ ,  for a given recorded configuration $\phi_r$ (the divergence between the rays after $\phi_r$ is exaggerated for illustration).   }
\end{center}\label{fig:screens}
\end{figure} 

Now, as in figure 5, suppose there are a sequence of such screens. 
A sequence of screens $\{\mathcal{S}^i_{(r)}\}_{i=1,2\cdots N}$ is a set of disjoint screens with the same record, and for which each element of  $\mathcal{S}^i_{(r)}$ is a record of some subset of $\mathcal{S}^j_{(r)}$ for $j>i$. Unfortunately, to prove conservation of probabilities, at this moment, we need to demand more; namely that there is no interference between extremal paths interpolating between the screens. For simplicity, we restrict the analysis to two screens and omit the subscript $(r)$, that tells us they refer to the same record. 

To calculate the  volume of the screen $\mathcal{S}_2$, we will use  the property of records, and the reciprocal relation of  the densities \eqref{equ:Van_vleck_volume} (valid in the no-interference limit) to relate the area-elements:\footnote{This reciprocal relation is heuristic: consider configuration space with Jacobi metric action, and a coordinate system parametrized by the outgoing geodesics, with the screens at constant distance. Then the geodesics are Eulerian, and their inverse density parametrizes the area element. To be more precise I know of no way other than to have a specific action functional. For a Jacobi system, with metric $G^{ab}$, the normal to the constant $S$ surfaces are given by $j_a=\Delta \diby{S}{\phi^a}=\Delta\nabla_a S$. Then reparametrization invariance implies that semi-classically one obtains the conservation equation $G^{ab}\nabla_a j_b =0$. For the full proof that this implies conservation, see \cite{QG_deco}. } 
$$\mathcal{D}(\phi^{(2)}_1)= |W( \phi_1^{(2)}, \phi_2)|^2\, \mathcal{D}(\phi_2)$$ 
where $\phi_1^{(2)}$ is the unique configuration in the first screen intersected by the extremal path between $\phi_r$ and $\phi_2$. 
Then the total area count of configurations  of the second screen is given by: 
\begin{eqnarray}
\nonumber\label{vol_screen2} A(\mathcal{S}_2)&=&|\psi(\phi_r)|^2\int_{\mathcal{S}_2} \mathcal D(\phi_2)  |W(\phi_r, \phi_2)|^2=|\psi(\phi_r)|^2\int_{\mathcal{S}_2} \mathcal D(\phi_2)  |W(\phi_r, \phi_1^{(2)})|^2 |W( \phi_1^{(2)}, \phi_2)|^2
\\ &=&
 |\psi(\phi_r)|^2\int_{\mathcal{S}_1} \mathcal D(\phi_1^{(2)})  |W(\phi_r, \phi^{(2)}_1)|^2= |\psi(\phi_r)|^2\int_{\mathcal{S}_1} \mathcal D(\phi_1)  |W(\phi_r, \phi_1)|^2=A(\mathcal{S}_1)
\end{eqnarray}
where,  assuming that the relation between the two screens given by the extremal paths was bijective, we removed reference to the second surface. 


\paragraph*{Sleeping Beauty}

In the (quantum) Sleeping Beauty paradox, an observer measures the x-spin of a spin-1/2 particle whose state is an eigenstate of z-spin. According to the many-worlds interpretation, the observer branches into two
successor observers, one of whom sees the result \lq{}spin up\rq{} and the other of whom sees the result
\lq{}spin down\rq{}. Let\rq{}s call the observer a \lq{}sleeping beauty\rq{}. On Sunday, she is  told the
following: we will put you to sleep and throw the  \lq{}spin coin\rq{}.  If the spin is up, we will wake you for an hour on Monday, put you back to sleep and wake you for an hour on Tuesday. When we put you back to sleep on Monday night, we will administer a drug that will cause you to forget the Monday waking. If the spin is down, we will wake you on Monday for an hour. And that is it.\rq\rq{}  Her room contains no indication of what day it is.   Each time she is asked a question: \lq\lq{}What is the probability you would assign that the original spin-coin came up or down?\rq\rq{} Figure 6 illustrates the conundrum. 

The debate revolves around one point: although she should always assign a probability of 1/2 to that type of event, here 2/3 of the times she was awakened, the spin was up. In the configuration space context explored here, there is a simple explanation for the disagreement, and it comes in precisely from the difference between records and screens. 

  \begin{figure}[h]
\begin{center}
\includegraphics[width=0.75\textwidth]{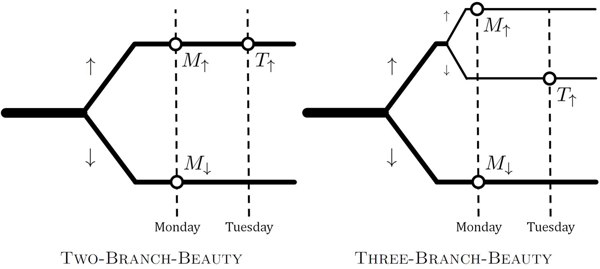}
\caption{The two setups: on the lhs the original one. On the rhs, a second setup, for which there is no difference between the conserved notion of probability and the way in which Beauty allocates her bets. The dotted lines can represent screens (their intersection with the waking up events is immaterial, for aesthetics purpose only).  (taken from \cite{Carroll}, creative commons license.)}
\end{center}\label{fig:beauty}
\end{figure}   

To describe this situation, we need four sorts of configurations: 1) the initial setup of the experiment -- let\rq{}s call this $\phi_o$ -- which is a record for the three subsequent ones. 2) Beauty waking up on Monday, with a recorded spin up  (outside of her sight). Let\rq{}s call this configuration $M_{\uparrow}$. 3)  Beauty waking up on Monday, with a recorded spin down  (outside of her sight). Let\rq{}s call this configuration $M_{\downarrow}$.  4)  Beauty waking up on Tuesday, with a recorded spin up  (outside of her sight). Let\rq{}s call this configuration $T_{\uparrow}$. 

The total volume for spin up is: $V(\uparrow)=V(M_{\uparrow})+V(T_{\uparrow})= 2V$, where we are assuming that each of these one hour events has a configuration space volume of $V$. The total volume for waking up with spin down is $V(\downarrow)=V(M_{\downarrow})=V$. If we want to ask the same question of screens, they don\rq{}t care about beauty waking up; all they care about is the area of the intersection with the semi-classical branches. In this case, we always have: 
$$A(S_1\cap \uparrow)=A=A(S_1\cap \downarrow)=A(S_2\cap \uparrow)=A(S_2\cap \downarrow).$$
where $A$ is the standard area element of the intersection. According to the volume measurements, we would have $V(\uparrow)=2V(\downarrow)$. Whereas for the screen intersecting areas we have e.g.: 
$$A(S_1\cap \uparrow)=A(S_1\cap \downarrow).$$
Thus the relative probability changes whether one considers just the volume of the events, or actually computes the area elements of the screens related to the two extremal histories of the experiment. 

The discrepancy could be seen to arise because here there is a redundancy of records, namely,  $T_{\uparrow}$ also has a record of  $M_{\uparrow}$. {Even though Beauty doesn\rq{}t know this,  she and her room  are significantly entangled with the rest of the environment (therefore it makes sense to speak of Tuesday and Monday inside the room as well).} Therefore counting this twice is overcounting. 

Therefore, if one wants to speak of conservation of probability, one should consider a screen, and thus assign the same probability for each of the spins. 
However, if one would expect Beauty to bet on her outcome, she should bet (with a 2/3 odds) that the spin was up, since she would collect this bet twice in the same \lq{}timeline\rq{}. 

The point being that although all of these conscious states of Sleeping Beauty  might be psychologically disconnected, the branches (the two alternative histories) still belong to the same string of records. 

  Note the difference between this case and another one with three alternatives. I.e. along \cite{Carroll},  instead of waking up Beauty twice when the spin is up, we instead observe another spin. If that second spin is also up, she is awakened on Monday, while if it is down, she is awakened on Tuesday. Again we ask what probability she would assign that the first spin was down. In this setting, there are no redundant records between the four configurations above, as can be seen clearly by following the extremal trajectories in the diagram.  In this case there is no controversy, and the way Beauty allocates her bets should follow these probabilities. 

 This example also touches on an argument relying on decision theory, used to find the Born probability from Many-Worlds \cite{Wallace_book}.  it is possible to drive a wedge between probability and betting (see Kent in \cite{Wallace_book}). 
 According to the Beauty\rq{}s best interest, she should bet at 2-1 on spin up, and yet this does not reflect the conserved probability.

\subsection{A note on Bayesian analysis and hypothesis-testing}\label{sec:Bayes}
Configurations with a record of multiple similar experiments  will increase or decrease our confidence regarding a given proposed distribution of the volume form $\mu(\phi)$. Note that here  the role of repeated experiments (i.e. many experiments in \textit{time}) is replaced by a notion of many records of an experiment at an instant (i.e. many similar experiments in space).

Let me first illustrate the issue  of hypothesis-testing in science,  and the meaning of the p-value, with the following standard example: Imagine you have a coin, which you would like to test against your hypothesis, namely, that the coin is fair. This is your \lq{}null hypothesis\rq{}. More generally,  it could be some prediction of your theory. You flip the coin 1000 times and get more heads than tails. The p-value won't tell you whether the coin is fair, but it will tell you the probability that you'd get at least as many heads as you did \textit{if the coin was fair}. If you got 900 heads,  the probability that your null hypothesis holds is small, but never zero, for it could have been that you were just extremely (un)lucky. 

Now,  think of all the configurations in $\mathcal{Q}$ that have the physical record of a thousand coin tosses, with all other things remaining the same. Given our volume form $\mu(\phi)$, we could partition such configurations into e.g. number of \lq{}heads\rq{}. Say region $A$ is a region with 400-600 heads, and region $B$ is a region with 1000 heads. According to your theory,  the volume of region $A$ is many times larger than the volume of region $B$, and yet, suppose you find yourself in region $B$. You cannot falsify the theory (in the sense that it now has no chance of yielding the correct volume form), because you still need to account for the chance that you are one of the few unlucky copies in region $B$. But the higher chances are that your initial volume-form was simply incorrect. The rule for performing this updating of your hypothesis is Baeysian analysis, which can be directly connected to the use of the Principal Principle \cite{Lewis}, described below.\footnote{Again, one shouldn\rq{}t count all of the instants in which one sees the coins for conservation of probabilities. Although in this case it is immaterial, because in all of the branches the behavior of the coin becomes static (wrt to some other subsystem, such as clock). }

The analysis takes into account that nothing can tell you the fact of the matter of the coin. Again, all you can know is the \lq\lq{}probability that you'd get at least as many heads as you did, given that your hypothesis is correct" -- which is the meaning of the p-value.  The more records of a given (approximately) reproducible phenomenon a set of configurations have, the more we test such null hypotheses. In this example, increasing the amount of coin tosses, the volume form $\mu(\phi)$ would sharpen around the predicted value of the theory, and any result outside of It -- say, all heads in the example above -- would cause you to update your confidence on your distribution.  Of course it would still be possible that your state is just on a fringe configuration, and your hypothesis is still correct. But this becomes more and more unlikely, forcing you to update your credence on your hypothesis.

The lack of certitude that necessarily remains  precisely mirrors that of single world interpretations of physics: one can never know if the next infinity of coin tosses will all be heads. One can only have varying levels of confidence in any given theory; these levels can be anything between very high or very very low, but never zero. It is easy to see that with enough people throwing coins, a small enough portion of these people could obtain any improbable result, and this small group might (mistakenly) conclude that the coin was unfair, whereas they were just \lq\lq{}unlucky\rq\rq{} (analogously, belonged to a fringe configuration in terms of configuration space volume). 

 While it is true that the theory is timeless, we can still give a meaning to active verbs such as \lq\lq{}updating\rq\rq{} (e.g. of our confidence level): to the extent that humans are  classical systems, extremal paths in configuration will reflect anything that the equations of motion predict, including rational (and irrational) \lq\lq{}updating\rq\rq{} of our theories.

\paragraph*{The \lq{}Principal Principle\rq{}}

To illustrate these thoughts, suppose I have a candidate volume-form representing my favorite theory, $\mu_{\mbox{\tiny fav}}$, defined by an action, by the field and symmetry contents and by $F$, given  above. Suppose that  $\mu_{\mbox{\tiny fav}}$ is immensely peaked in a given region. Suppose further that I found myself outside of this region. Then the confidence level I originally had on  $\mu_{\mbox{\tiny fav}}$  decreases. This level, however, is never absolutely zero, for there is always a chance that my original volume-form is correct and yet I am one of the (proportionally) few unlucky 'copies' of myself that lies outside of the peaked region. 

Updating of knowledge occurs solely as a deterministic system. The volume-form concentrates on classical trajectories, and these encode the evolution of rational mechanisms such as ourselves. 
The role of science in this context is to discover the appropriate configuration space,  and its volume-form; which is predicated on discovering its instantaneous symmetry group (and the most homogeneous element under its action) and the appropriate action functional.
The meaning of \lq\lq{}degrees of belief\rq\rq{} and \lq\lq{}chance\rq\rq{} are in complete accord with those outlined by \lq\lq{}The Principal Principle\rq\rq{} \cite{Lewis} (and thus able to reproduce all properties of the usual notions), as I now explain.

The description of chance, degree of credence, and updating of such a degree are enough for any use of the theory of probabilities in the real world. This point of view, made precise in the seminal paper by Lewis \cite{Lewis}, will be here very briefly recounted and adapted to our setting. A complete analysis of the analogy, and in particular its transposition to the timeless context, will be left for a further paper. 

According to \cite{Lewis}, \lq\lq{}Assume we have a number $x$, proposition $A$, time $t$, rational agent whose evidence is entirely about times up to and including $t$, and a proposition $E$ that is about times up to and including $t$ and $X_{x,A}$  entails that the chance of $A$ at $t$ is $x$. In any such case, the agent's credence in $A$ given $E$ and $X$, is $x$.\rq\rq{} The principle then connects information about chances to rational credences, via the formula $C(A/X_{x,A}E)=x$. 

In the translation to our case, $E$ is  a  set in configuration space characterized by having the same set of records, e.g. $E\subset \mathcal{Q}_{(r)}$. To make matters more easily translatable, we will also assume here that $E$ is in fact a screen, so $E=\mathcal{S}_{(r)}$. Then  $C(-/E)$, the initial credence function conditionalized on one\rq{}s present total evidence, $E$, will be the volume of the screen with respect to an observer\rq{}s  choice of volume-form $\mu(\phi)$.\footnote{In the work here, made implicitly through a choice of action functional, initial point, coarse-graining, etc.} A further sub-region of $E$, which is the proposition one wants to assign a chance to, is called $A$. Finally, $X_{x,A}$,  the proposition that the chance of $A$ holding at $t$ is $x$, gives the total credence function $C(A/X_{x,A}E)=x$, which obeys, for the total credence on $A$:
$$C(A/E)=\sum _x C(X_{x,A}/E)x$$
where $C(X_{x,A}/E)$ is the certainty one has that the probability for $A$ given $E$ is $x$.

  In our case,  the nonnegative, finitely additive volume-form $\mu(\phi)$, gives us a credence function, through $C(A/E)=V(A)/V(E)$, i.e. the relative volume, according to a given $\mu(\phi)$, of the region $A$ wrt to the total region compatible with the evidence $E$.  Similarly, $C(X_{x,A}/E)$ further partitions $A$ into sub-regions of relative volume $x$. 
  
  In our case, it is easy to dismiss  \lq{}knowledge of future events\rq{}, for these are merely events that are not records of the region, and are thus not in $E$. Moreover, the result that \lq\lq{}the past is no longer chancy\rq\rq{}, $1=C(A/X_{x,A}A)=x$, here is made trivial: it refers to a record within its own screen, which is one. 
  
    Notably,  in this way, one does not really invoke frequentism: \lq\lq{}it makes sense to speak of the chance of getting exactly seven heads on a particular occasion; equally it makes sense speak of the chance of getting exactly seven heads on a particular sequence of eleven tosses.\rq\rq{} \cite{Lewis}

\section{Impact on the interpretation of quantum mechanics}\label{sec:conclusions}

As has been said before, the main problem with many worlds is that none of us wants to think that ``I am a mere copy of myself". 
In his contribution to \cite{Wallace_book}, Wallace writes  ``According to our best current physics, branches are real", and this is roughly the position I take in this paper as well.  As I explained, my interpretation is not exactly of branches -- there is no instant in which branching takes place -- but there is a sense in which the ontological existence of configuration space  is my version of the many worlds scenario. To be more precise, nothing is splitting, all the individual copies of the system already exist in timeless $\mathcal{Q}$. The set of all configurations exist, with a volume form on top of it, counting regions with similar configurations. 

There is no measurement process, only records embedded in configurations and a relative volume of said configurations. This volume can be proven  to be given by a Born norm. Crucially, here the Born rule is not associated to a measurement, and there is no definite-outcome problem,  as in standard Many-Worlds.

These are the  arguments.  I will now slightly expand on these points and on the impact this view might have on foundational problems of quantum mechanics.

\subsubsection*{Classicality and decoherence}

 In the present  formulation the problem faced by decoherence is no longer to account for the observable classical reality from the basis of a quantum theory: each configuration is, tautologically, classical. The lack of observed superpositions here \emph{precedes decoherence}. Empirically, it is only through interference effects that we \emph{can} observe superpositions of states in the first place.
 Decoherence is still indeed the explanation for the absence of interference, but it is not necessary in order to abolish the existence of macroscopic (or microscopic) superpositions. 
 
Now, does the present work offer evidence that in this context decoherence alone solves the measurement problem? Let us try to address this by using two representative counter-arguments against the usual claim in this direction. First, Joos in  \cite{Joos}: 
\begin{quote}
Does decoherence solve the measurement problem?
Clearly not. What decoherence tells us, is
that certain objects appear classical when they
are observed. But what is an observation? At
some stage, we still have to apply the usual probability
rules of quantum theory.
\end{quote}
In his review \cite{Laloe}, Laloe states \begin{quote}Indeed, the emergence of a single result in
a single experiment, in other words the disappearance of macroscopic superpositions,
is a major issue; the fact that such superpositions cannot be
resolved at any stage within the linear Schr\"odinger equation may be seen
as the major difficulty of quantum mechanics. As Pearle nicely expresses it, the problem is to \lq\lq{}explain why events occur!\rq\rq{}\end{quote}
  The present work, goes very much against the modern canon that using classical concepts in order to describe quantum mechanics is ill-conceived. By positing first the existence of a classical configuration space, it straightforwardly answers Joos, Laloe and Pearle: i) there are no macroscopic superpositions at any point of configuration space, superpositions are interference effects between different histories of the system, and ii) events, or ``instants" of the classical Universe are all that exists, there is no need to explain them.  In the present context, the answer to the question of definite outcomes is unambiguous: it is the ontological existence of configuration space -- with each point including the instantaneous brain state of any possible observer -- that resolves it. 

\subsubsection*{The Born rule: avoiding criticisms applicable to Many Worlds}   

In section \ref{sec:Born}, I have shown that the Born rule emerges as a way to \lq{}count\rq{} configurations in $\mathcal{Q}$. In fact, it emerges from  the factorization property of records and  reduction to the `purely classical' density (i.e. the one that would  result from a purely classical propagation of the density of a region around $\phi^*$, without interference).

 %

Irrespective of this uniqueness, I believe the current framework avoids major criticisms to Many Worlds. The usual criticism of probabilities in the Many Worlds context is that no matter what probability distribution your theory ascribes to events, there are always branches that will confirm the theory, and therefore the proposed distribution cannot be falsified. This is also the case here. All science can do through repeated experimental tests in an uncertain world is to diminish the chance that we are confirming a false theory. We can only make a false-positive region smaller and smaller and hope (with better and better odds!) that we are not in it. This is as much true for medical science and ordinary coin tosses 
 as it is for a density distribution in configuration space.

Moreover, here \textit{I can implement a Born rule without requiring the notion of a non-unitary measurement}. The Born rule is merely a particular way of counting instantaneous configurations; it does not need to arise from the dynamics itself.  

This avoids a standard criticism to Hartle\rq{}s argument \cite{Hartle_68}. 
The criticism is that  one requires \emph{the Born rule itself} to measure the relevance of the deviation between the action of frequency operators  and the  square of the eigenvalues associated to a measurement \emph{by Born\rq{}s rule}. For the present approach this debate is   moot: the key point is, again, that unlike in Many Worlds, we here do not need to \lq{}explain why events occur\rq{}, in the words of Pearle (as quoted by Laloe \cite{Laloe}). The set of all configurations exists, and on top of it, a volume form, dictated by an action functional and a uniquely specified boundary condition. {I should also note that, as made explicit in \cite{QG_deco}, this boundary condition is unique only for certain field content, instantaneous symmetry content, and spatial topologies.} 


  I have crucially assumed that the volume form over configuration space is a \lq\lq{}factorizable\rq\rq{}  measure based on the quantum-mechanical transition amplitude. From the mathematical perspective, this guarantees that when $\Psi(\phi)=\Psi_1(\phi)\Psi_2(\phi)$ then the measure also follows suit.  The perhaps more physical justification  is that it is necessary for a type of Markovian property. Namely, suppose that there are $n$  configurations $\phi_i$ with the same record, $\phi_r$. We can think of $\phi_i$ as possible outcomes of a given laboratory experiment ---say two different Geiger detectors clicking  --- and $\phi_r$ as the entire setup of the experiment --- the preparation of an electron emission gun, but without activation of the detectors. Then, we have $\Psi(\phi_i)=W(\phi^*,\phi_r)W(\phi_r, \phi_i)$. The ratio of amplitudes is 
$$\frac{\Psi(\phi_i)}{\Psi(\phi_j)}=\frac{W(\phi_r, \phi_i)}{W(\phi_r, \phi_i)}$$
 It thus \lq\lq{}forgets\rq\rq{}  the amplitude of the setup itself. Allowing probabilities to follow suit  is what enables one to set up initial conditions for experiments, with no regard to the history of the Universe prior to that moment. 
   
   The emergence of the Born rule is not really that surprising; what has been achieved is basically just a re-interpretation of the source of the Born rule in a probabilistic measure that avoids some of the interpretational problems referenced here. 

It is also important to note that even if the volumes of observers can be formally infinite, and its ratios not straightforwardly definable,  the use of records and the factorization property makes it easier to cancel out identical infinities and obtain meaningful ratios. This cancellation only becomes really important once we take into account issues of locality. In the accompanying paper \cite{Locality_riem}, it is shown how the amplitude kernel also factorizes for dynamically disconnected regions of physical space, and thus at the very least we can say that when our system has been localized to a finite-dimensional configuration space, the ratios are well-defined. 

That is, suppose we have access to a given region of space, or some coarse-grained configuration variables.  Universality and decomposition arguments, together with the locality arguments of  \cite{Locality_riem}, then imply that the volume form decomposes as well. More specifically, suppose that for the region $R\subset \mathcal{Q}$, we have a decomposition, $\mathcal{Q}_{|R}= \tilde{\mathcal{Q}}\times \mathcal{Q\rq{}}$ and $\mu(\phi)=\mu(\tilde\phi)\mu(\phi\rq{})$, where $\tilde\phi\in \tilde{\mathcal{Q}}\subset \mathcal{Q}$, and $\phi\rq{} \in \mathcal{Q\rq{}}\subset \mathcal{Q}$.  Suppose  all we have access to are the partial configuration variables $\tilde\phi$. 
Then, even if the volume of the manifold $$\mathcal{Q}_{\tilde\phi_1}:=\{\phi\in \mathcal{Q}\,~ |\,~ \mbox{pr}_{\tilde\phi}(\phi)=\tilde\phi_1\}$$  will be usually infinite,  the ratios of volumes for two different partial configuration values:
$$\frac{\int_{\mathcal{Q}_{\tilde\phi_1}} [\mathcal{D}\phi] \mu(\phi)}{\int_{\mathcal{Q}_{\tilde\phi_2}} [\mathcal{D}\phi] \mu(\phi)}=\frac{\int [\mathcal{D}\tilde\phi] \mu(\tilde\phi)\delta(\tilde\phi_1)\int [\mathcal{D}\phi\rq{}] \mu(\phi\rq{})}{\int [\mathcal{D}\tilde\phi] \mu(\tilde\phi)\delta(\tilde\phi_2)\int [\mathcal{D}\phi\rq{}] \mu(\phi\rq{})}=\frac{\mu(\tilde\phi_1)}{\mu(\tilde\phi_2)}$$
can be finite. 

This approach provides a different take on regularization procedures; one does not \lq\lq{}brush infinities under the rug\rq\rq{} but takes quotients of similarly defined infinities.   Nonetheless, in practice this quotienting is technically similar to e.g. a minimal subtraction scheme --- a scheme used to absorb the infinities that arise in perturbative calculations beyond leading order, consisting of absorbing only the divergent part of the radiative corrections into the counterterms.
  In the analogy, the regulator identifies the degrees of freedom that differ between the two regions, and the quotient removes the identical infinities associated to the infinite-volume of the ultraviolet degrees of freedom. This analogy will be elaborated upon in further work.

\subsubsection*{The basis selection problem and configuration space.} 

Given a `system + environment' decomposition, decoherence is supposed to choose a pointer basis for the system, the basis onto which the reduced density matrix diagonalizes. Is our choice of configuration space aligned with this criterion?  

 The usual criterion for selecting a pointer basis for a system is based on the interaction with the environment, and embodies the idea of robustness of correlations. The selection is thus a property of the interaction Hamiltonian, which acts by determining what states lead to stable, perceivable records when the interaction of the system with the environment is taken into account.
The fact that the pointer basis is in most cases the position basis is seen as an effect of the manner by which we write local interactions. 
 
 Here, I \emph{start} with the assumption of an underlying real existence of configuration space. Then the fact that the preferred basis is usually given in the position basis is seen by me as very convenient for the formulation in configuration space attempted in this paper.  Furthermore, the only  possible issues with taking the position basis as the pointer basis of decoherence occur for relativistic theories. As Wallace points out \cite{Wallace_role} (my emphasis): 
\begin{quote}
In the case of non-relativistic quantum theory [the selection of the preferred basis] is unproblematic. The decoherence-preferred basis is basically a coarse-graining of the position basis, so a collapse rule that collapses the wave-function onto wavepackets fairly concentrated  around a particular center of mass position, or a choice of position as the  hidden-variable, will do nicely. [...] It is crucial to note what makes this possible. Position has a dual role in
non-relativistic quantum theory: it is at one and the same time (a) one of the
fundamental microphysical variables in terms of which the theory is defined,
and (b) such that a coarse-grained version of it is preferred by the high-level,
dynamical, emergent process of decoherence. \emph{As such, it is possible to formulate
modifications or supplements to non-relativistic quantum theory that are
both precisely defined in terms of the microphysical variables used to formulate
quantum mechanics, and appropriately aligned with the macrophysical variables
picked out by decoherence.} 
\end{quote}
Here no issue can thus arise with choosing an ontological status for the position basis. The present model interprets reasons (a) and (b) given above differently from  Wallace however. In this interpretation, (a) and (b) are emergent from the primary existence of a kinematically non-relativistic theory in configuration space. 
   
 However, standard decoherence still demands a decomposition into `system + environment', which can be quite problematic,  not only in the absence of subsystems. Issues with uniqueness of the split can also arise,  with many  splits yielding a diagonalization of the reduced density matrix. 
 
 Consistent histories is a framework which can avoid the specification of this split. In consistent histories the decoherence functional describes the orthogonality of branches of the wavefunction attached to each coarse-graining element.  
 
  Nonetheless, in  consistent histories, superstructures that need to be put in by hand and are not specified by the theory still linger in the form of choices of `single frameworks'.   In the present context we have constructions very similar to consistent histories, in that they do not necessarily involve a separation between system and environment. However, unlike what is the case in consistent histories, we have a  natural definition of framework in the existence of configuration space and   preferred coarse-grainings within that. 

 \paragraph*{The set selection problem and preferred coarse-grainings}

 In the canonical setting of consistent histories, the crucial criteria for a set of projectors to yield probabilities is for each ``branch" to be orthogonal to the others, and for the branches to form a decomposition of unity, the so-called `consistency and completeness conditions'. Even so, there are uncountably infinite ways to choose complete, consistent sets of projectors (which may be non-commuting, and thus inconsistent\footnote{Projectors of a single history commute, but associated to different choices they might not: in QM there is no truth functional associated to the product of non-commuting projectors: $PQ$, even if $Q$ and $P$ are projectors (associated to truth functionals). One can only combine different frameworks when all projectors commute (in which case there is a larger set for which both subsets are fine-grainings). Once one has a certain splitting of the entire system under consideration into an `environment+system+apparatus' tensorial product, one can use (generalizations of) the tri-decomposition uniqueness theorem, which states that for an orthogonal basis for each subsystem there is a unique choice of branches, but of course this begs the question of where the separation itself comes from, which is what consistent histories was supposed to evade.}) and  no objective principle to guide this choice.
 
 Measurements in different `realms' or `frameworks' most often  occur in different regions of configuration space, and thus have processes represented  by completely different coarse-grainings. In essence, we are not projecting some underlying quantum state of the Universe onto a given basis, we are measuring the relative frequency of configurations with the same memories (records). This is what makes the framework contextual in the sense of experiments, and non-contextual in the sense that a global state exists.

  In the present context, even allowing for the possibility that using configuration space as a preferred basis still leaves open  choices for coarse-grainings of paths in $\cal Q$, I proposed a method to select preferred ones,  the extremal coarse-grainings (ECs), (see definition \ref{def:ECS}). Or, in a more general  (and conjectural) context, the minimal piece-wise extremal ones -- minimal PECs (see appendix \ref{app:piecewise}).\footnote{One should note that Fujiwara's et al \cite{Fujiwara} formalization of the path integral through piece-wise classical paths is one of the more robust formal mathematical treatments of the path integral. } The benefit of these choices is that they avoid a valid objection to a usual choice of coarse-grainings in the path integral context. The choice is that of all paths crossing a certain region in space-time (in our case, configuration space). The objection, due to Halliwell and collaborators, is that one can only do that by effectively implementing reflective boundary conditions on the region, which then becomes subject to the quantum Zeno effect \cite{Halliwell_Zeno}.  
 
   In the present case, using the Riemannian exponential map around an extremal path, one can build a coarse-graining without implementing such conditions, for which paths in a given element of the coarse-graining also never leave that element. 
    The  radius of the ECs defines the coarseness of the sets, or the `distinguishibility of states' in terms of the accuracy of our amplitude kernel. The point is that the radial length has an intimate relation with the decay of  the amplitude kernel for generic systems. Indistinguishibility here means that coarse-grainings that (resp. don't)  decohere  to the given order for one given configuration, will also (resp. not) decohere for the `indistinguishable' configuration; meaning that to a certain approximation one would (resp. wouldn\rq{}t) observe interference.

\section*{ACKNOWLEDGEMENTS}

I  would like to thank Lee Smolin, Flavio Mercati, Tim Koslowski, and Simon Saunders for comments, and Clement Delcamp for help with the figures. This research was supported  by Perimeter Institute for Theoretical Physics. Research at Perimeter Institute is supported by the Government of Canada through Industry Canada and by the Province of Ontario through the Ministry of Research and Innovation.

\begin{appendix}

 \section*{APPENDIX}

\section{The Jacobi metric}\label{app:Jacobi_metric}

The Jacobi version of the Maupertuis principle establishes some instances when  dynamics
 can be viewed as geodesic motion in an associated Riemannian
manifold. That is, if the action can be written as
$$ S = \int (T - U)dt$$
for a system defined in a Riemannian manifold, $(M, g)$, with {\emph smooth} potential $U$, the extremal trajectories
of $S$ with energy $E = T + U$ coincide with the extremals (geodesics) of the
length functional 
\be\label{equ:jacobi_length}L[\gamma]=\int |\gamma|ds
\ee
 defined in $(M, h)$, where $h$ is the constructed Jacobi metric, conformally related to $g$ by $h = 2(E - V)g$, for $E>V$, and the norm is calculated using this metric.\footnote{Note that for $E<V$ we have an imaginary length functional. Accordingly, tunneling emerges from paths that extremize the Euclidean action, or seen otherwise, as paths that are themselves imaginary. We do not want to expand on this subject here. For more information on this \cite{Turok_real_time}, and on how to obtain the correct tunneling amplitude from a real path integral, see \cite{Tanizaki}. }
In general I will call \emph{a Jacobi metric}, any  metric in configuration space whose geodesics are extremal paths of the action.

\subsection*{Geodesically and dynamically connected manifolds}
For any finite-dimensional metrically complete Riemannian manifold, the Hopf-Rinow theorem  guarantees that two points of the manifold will be connected by a geodesic.\footnote{ For Finsler metrics and infinite-dimensions there are caveats (the geodesics come arbitrarily close to any given endpoint in infinite-dimensions for example), but nothing that would bother us here.}

However, even for finite-dimensional spaces, there is already here a very robust obstruction to this procedure: namely, non-smooth potentials $U$ (such as a Dirac $\delta$-function potential, which are ubiquitous in physics). In that case, it can be shown that there might be no extremal paths connecting two given points $a$ and $b$. However, Marsden has shown that by considering geodesic flows with corners, one can re-obtain a version of the Hopf-Rinow theorem for systems with a Jacobi metric. That is, one must use piece-wise geodesics.  For questions on the generic behavior of ``critical connectedness" in Hamiltonian systems, see \cite{Marsden}.

\section{DeWitt expansion and piecewise extremal paths.}\label{app:dewitt}

 \subsection{The DeWitt semi-classical expansion}\label{sec:dewitt}

\paragraph*{Coarse-grainings and Zeno}

The simplest type of coarse-graining one could try to define would be through paths which never leave a certain region  $O\subset \mathcal{Q}$.  However, as shown in \cite{Halliwell_Zeno}, this definition will suffer from the quantum Zeno effect. This happens as follows: In standard particle quantum mechanics,  upon calculating the path integral by the limit:
\be\label{equ:zeno} W(\mathbf{x}_i,t_f;\mathbf{x}_f,t_f)=\lim_{ n\rightarrow \infty} \int_R d^dx^1\cdots \int_R d^dx^{n-1} W(\mathbf{x}_k,t_k;\mathbf{x}_{k-1},t_{k-1})\ee
where $R$ is the space-time region, each integration acts as a projector of the state, by 
$$P=\int_R d^d x|x\rangle\langle x|$$
and in the  limit, this constitutes a continuous projector of the state onto the region. By the quantum zeno effect, this implies that the restricted propagation is already unitary, or, equivalently, the particle is  never allowed to leave $R$. Equivalently, Halliwell has shown that defining a set of paths by \lq\lq{}never entering\rq\rq{} a region, amounts to implementing reflecting boundary conditions  on said region, which alter the path integral one would like to study. 

Thus the issue to evade with our choice of coarse-graininings in configuration space is the formalization of a coarse-graining implicitly through reflecting boundary conditions. To do so,  I will use a different approach, which is based on the techniques developed in \cite{Cecille} for the computation of the path integral based on integrating over deviation vector fields at geodesics. That work is too technical for a comprehensive treatment here. The important point for us, is only that, if an extremal path between $\mathbf{x}_i$ and $\mathbf{x}_f$ exists, call it $\gamma_{\mbox{\tiny{cl}}}$, one can replace the integration over paths by an integration over the space of deviation vector fields over $\gamma$, with a measure given by the Jacobian matrix.
 The benefit is that this will give a constructive definition for coarse-grainings; not an implicit one as e.g. \emph{all paths} that enter or don\rq{}t enter a region (which gives rise to the quantum zeno effect).

Here I will very briefly report on some results of Cecille deWitt in \cite{Cecille}, concerning a higher order semi-classical expansion around extremal paths. In that paper, the problem of computing the path integral is re-expressed completely in terms of classical paths and variations around it. The rationale is to expand the action in higher variations around a segmented classical path, obtaining arbitrarily high orders of approximation to the full path integral. Although the case studied in \cite{Cecille} is for finite-dimensional manifolds, the work was extended to infinite-dimensional Banach spaces in \cite{Clarke}.

The technical steps that make it possible are, very roughly,
\begin{enumerate} 
\item The path integration can be replaced by an integral over the space of vector fields $\mathbb{X}$ (along the stationary path) which vanish at the endpoints. This is still assumed for all possible vector fields, therefore it encompasses all the paths. 
\item One segments the extremal path into many subsegments. Now let $\beta(u,t)$ be a geodesic variation of $\gamma_{\mbox{\tiny{cl}}}(t)$, i.e. one  which does not necessarily keep the endpoints fixed and is itself extremal, i.e.  $\frac{d}{du}_{|u=0}S[\beta(u)]=0$ and $\beta(0)=\gamma_{\mbox{\tiny{cl}}}$.  A Jacobi vector field, $\beta\rq{}(0)$,  can be defined from initial and final data on each segment.  Using broken Jacobi vector fields at each segment, for geodesic variations that vanish at one endpoint but not the other of each of these segments, one can find a measure on the space of deviation vector fields  $\mathcal{D}(u)$, which has cylindrical (i.e. composition) properties. 
\item  This measure involves an integration over Jacobi fields at each endpoint (not at midpoints of the segments). This is because the Jacobi matrix serves as an analogue of the Feynman propagator. It is a matrix relating initial and final vectors.
\item It is then shown that the infinite limit of the path integral obtained with this projected measure at each segment yields the standard path integral. 
\be \label{equ:higher_kernel1}
W({\phi^*}, \phi_f)=\Delta_\gamma^{1/2}e^{iS[\gamma]/\hbar}\sum_{n=0}^\infty (i\hbar)^n A_n\ee 
where $A_0=1$ and the next orders depend on further variations around the classical action (and their projection to the cylindrical measures at the endpoints through the  Jacobi matrix). 
\end{enumerate}
Equation \eqref{equ:higher_kernel1}   may provide a way to maintain preferred coarse-grainings  centered on extremal paths and yet obtain higher order of expansions in $\hbar$ beyond the usual semi-classical approximation. 
\footnote{ The formal treatment of the time-slicing approach used here in the context of piece-wise extremal curves was developed in \cite{Fujiwara}.}

\subsection{The piecewise semi-classical approximation}\label{app:PEC}
There might be cases in which there are no extremal paths between two given configurations ${\phi^*}$ and $\phi_f$. This failure can be most easily visualized by connecting the least action principle to the existence of a corresponding metric and geodesics over the same space, in which case the manifold with the induced metric becomes smooth-geodesically incomplete, as discussed at length in \cite{Marsden}, and briefly recounted in appendix \ref{app:Jacobi_metric}. {  The easiest way in which we can envisage this property even in the finite-dimensional case are for non-smooth potentials (such as Dirac delta barriers, or mirrors) in configuration space whose associated symplectic flows have ``corners"   \cite{Marsden} (or here called vertices).}  

 On the other hand, even for non-geodesically complete Riemannian manifolds, if two of its points  are in the same connected component  they can  be connected by paths which piece-wise are geodesics. In the metric case, the space of all paths on $\cal M$  can be densely covered by the (inductive limit in $N$ of the) space of piecewise geodesic segments: 
\be\label{equ:piecewise}\Gamma^N_{\mbox{\tiny pc.geod}}({\phi^*},\phi_f)=\{\prod_{i=0}^{N}\gamma_i~|~\gamma_i:[0,1]\rightarrow\mathcal{Q}~ \mbox{is geodesic, and}~~\gamma_{i+1}(0)=\gamma_i(1),  \forall i, \gamma_0(0)={\phi^*}, \gamma_N(1)=\phi\}\ee
Let us assume that this is also the case here -- mutatis mutandi for  configuration space (endowed with a Lagrangian density) instead of a Riemannian manifold, and extremal paths wrt the action we have on configuration space instead of geodesics. 

The piece-wise approximation used here can be roughly seen as the inverse procedure as the Feynman - Dirac standard one of obtaining the path integral, using the decomposition of unity at successive time intervals. This approach to the Feynman path integral is known as the ``time slicing approximation with piece-wise classical paths", and it is extensively used in the literature not only for a formal definition of the path integral (see [Fujiwara] \cite{Fujiwara} for a rigorous treatment), but also as a practical method of computation.  In the simplest case, it consists in splitting the usual particle path integral  time interval into regular $N+1$ sub-intervals $\delta t$, the amplitude kernel becomes
\be\label{equ:particle_segmented} W((q_i,0), (q_f, \tau))= \int\prod_{i=1}^N\langle (q_f,\tau)|(q_N,N\delta \tau)\rangle\cdots\langle (q_1,\delta\tau)|(q_i,0)\rangle
\ee
One obtains the path integral by taking the infinite limit $N\rightarrow\infty$. The proposal is to take the time intervals small enough so that one can always find a classical path between the intermediary points  and approximate each segment semi-classically.  

Here the proposal is precisely parallel to the particle quantum mechanics case. The main difference is that in the absence of time we take the Jacobi-length intervals of the action to be small, and since we have not rigorously proven the analogous result, we state it as a conjecture: 
\begin{conj}[The piece-wise expansion]\label{conjecture}
Given any two configurations ${\phi^*}$ and $\phi_f$ (not necessarily connected by an extremal path, in an action of Jacobi type), for a given order $\epsilon$, there is a number $N(\epsilon)$ and a finite distance $D(\epsilon)>0$ (in the Jacobi metric) such that 
\be\label{equ:conjecture} W({\phi^*},\phi_f)= \int \prod_{j=1}^N{\cal D}\phi_j W({\phi^*}, \phi_1)\cdots W(\phi_N, \phi_f) +\order\epsilon
\ee where extremal paths of length $D/N$ exist between $\phi_j$ and $\phi_{j+1}$ (and thus $W(\phi_j, \phi_{j+1})$ is given to any order in $\hbar$  by equation \eqref{equ:higher_kernel1},  and to first order in $\hbar$ by \eqref{equ:semi_classical_exp}). 
\end{conj}
If there is a unique extremal path between a given  $\phi_{j-1}$ and $\phi_{j+1}$, one can use equation \eqref{equ:semi_classical_record} to ensure that $\phi_j$ is in that segment (thereby collapsing this subdivision). The approximation holds true for $\epsilon\sim \hbar$ in the trivial case when ${\phi^*}$ and $\phi_f$ are classically connected, by equation \eqref{equ:higher_kernel1}, when $N(\epsilon)=1$ and $D(\epsilon)$ is the geodesic distance between the points. It also holds for  $\epsilon=0$  in the limit $N\rightarrow \infty$ (as then we can approximate each segment by the Lagrangian, as in the infinite limit of \eqref{equ:particle_segmented}). 

For instance, if we focus on the space of piecewise geodesics with two segments, between $v_{i-1}$ and $v_{i+1}$, one has a well-defined calculus of variations where the position of the intermediary vertex $v_i$ is the variation parameter, and it makes sense to say that a certain $v_i^*$ is extremal. A simple, and yet non-trivial example in 2D Euclidean geometry is to consider two points (without loss of generality), $p_1=(x_1,0)$ and $p_2=(x_2,0)$ and remove from $\mathbb{R}^2$ an open line segment $L=(0,]-a,a[)$ between $p_3=(0,-a)$ and $p_4=(0,a)$, so as to make the manifold geodesically incomplete. The variational principle yields the two piecewise straight segments $\gamma_1=\overline{p_1p_3}\cdot \overline{p_3p_2} $  and $\gamma_2=\overline{p_1p_4}\cdot \overline{p_4p_2}$.

\subsection{Piecewise extremal coarse-grainings}\label{app:piecewise}
Using definition  \ref{def:ECS} for EC\rq{}s, we define
\begin{defi}[Piecewise extremal coarse-grainings (PECs)]\label{def:PEC}
A piecewise extremal coarse graining (PEC) is a  coarse-graining, where each element of the coarse graining set, $C_{\alpha}$, is composed of extremal ``segments":
\be\label{equ:piecewise_coarse}
\{C_\alpha=C_{\gamma^\alpha_1}\cdot C_{\gamma^\alpha_2}  \dots C_{\gamma^\alpha_{N_\alpha}}~|~\gamma^\alpha_1\cdot\gamma^\alpha_2\dots\gamma^\alpha_{N_\alpha}\in \Gamma^{N_\alpha}({\phi^*},\phi_f), \alpha\in I\}
\ee
where each bundle of paths is defined as in the previous  definition \ref{def:ECS} as contained in its tubular neighborhood $\nu_{\gamma^\alpha_i}(\rho_\alpha)$. 
The intermediary points (the endpoints of each segment) will be called the \emph{vertices} of the PEC. \end{defi}

\paragraph*{Set selection and PECS}

 Let me first define the order of a PEC  as the largest number of segments of any element of the PEC. From \eqref{equ:piecewise_coarse}: 
\be\label{equ:order}
\order I:=\max_{\alpha\in I}{N_\alpha}
\ee  A minimal-$\epsilon$ PEC is one that has the minimal order and is still exhaustive to order $\epsilon$:
\be\label{equ:minimal}
\{C_\alpha\in \Gamma^{N_\alpha}({\phi^*},\phi_f), \alpha\in I_{{\mbox\tiny min}}~|~\order{I_{{\mbox\tiny min}}}= \inf_{I\in \mbox{\tiny{PEC}}({\phi^*},\phi_f)}{\order I}\}
\ee  
where $\mbox{PEC}({\phi^*},\phi_f)$ parametrizes the space of all possible piecewise extremal coarse-grainings between the two configurations.  
Apart from the formalities, a minimal PEC, if it exists, can be seen as just that minimal set of segments of extremal paths such that its corresponding PEC yields the approximate amplitude for the given process. In that way, I propose that the existence of a  unique minimal PEC yields the \emph{preferred coarse-graining} for any given transition amplitude between an initial and a final configuration. 

Uniqueness of the minimal PEC should be related to a unique solution of the  variational problem of the transition amplitude wrt the vertices of the minimal PEC.
A non-trivial example of a minimal PEC is of course the semi-classical approximation when extremal paths (without segmentation) exist between ${\phi^*}$ and $\phi$, (and $\phi$ is not a focusing point). In this case it is given implicitly by \eqref{equ:semi_classical_exp}, i.e. each extremal path seeds one of the coarse-grained sets (and indeed variation on an  intermediary vertex will project it back to the extremal curve).   However, this is the only case in which we use minimal PECs quantitavely in this paper.   A more quantitative treatment of this approach using the expansion \eqref{equ:higher_kernel1} is possible but will not be explored here..

Using minimal PECs, I can expand the definition of the semi-classical record used in definition \ref{def:full_record}, 
 \begin{defi}[Semi-classical record (general)]\label{def:record_PEC}
Given an initial configuration ${\phi^*}$, $\phi$ and  $\{C_\alpha\}_{\alpha\in I}$ the minimal $\epsilon$-extremal coarse graining (see def. \ref{def:PEC} for PECs and equation \eqref{equ:minimal} for the minimal criterion) between $\phi^*$ and $\phi$,  of radius $\rho_{\mbox{\tiny max}}$, $\phi$ holds  a semi-classical  record of a field configuration $\phi_r$ to order $\epsilon$, if the ball $B_{\rho_{\mbox{\tiny min}}}(\phi_r)$ is contained in every $C_\alpha$, i.e.   $B_{\rho_{\mbox{\tiny min}}}(\phi_r)\subset C_\alpha, \forall \alpha\in I$.  
\end{defi}

     \end{appendix}



\end{document}